\documentclass[aps,prb,preprint,showpacs,showkeys,tightenlines,
preprintnumbers]{revtex4}     
\usepackage{bm}
\usepackage{graphicx}

\begin{document}

\hfill In press, PRB.
\title{Nonequilibrium Green's function theory for transport
and gain properties of quantum cascade structures}
\author{S.-C.~Lee}
\author{A.~Wacker}
\affiliation{Institut f\"{u}r Theoretische Physik, 
Technische Universit\"{a}t Berlin, Hardenbergstra{\ss}e 36, 10623 Berlin,
Germany}

\date{\today}

\begin{abstract}

The transport and gain properties of quantum cascade (QC) structures
are investigated using a nonequilibrium Green's function (NGF) theory
which includes quantum effects beyond a Boltzmann transport description.
In the NGF theory, we include
interface roughness, impurity, and electron-phonon scattering
processes within a self-consistent Born approximation,
and electron-electron scattering in a mean-field approximation.
With this theory we obtain a description of the nonequilibrium
stationary state of QC structures under an applied bias,
and hence we determine transport properties, such as the current-voltage
characteristic of these structures. We define two contributions
to the current, one contribution driven by the scattering-free
part of the Hamiltonian, and the other driven by the scattering
Hamiltonian. We find that the dominant part of the current
in these structures, in contrast to simple superlattice
structures, is governed mainly by the scattering Hamiltonian.
In addition, by considering the linear response of the
stationary state of the structure to an applied optical field,
we determine the linear susceptibility, and
hence the gain or absorption spectra of the structure.
A comparison of the spectra obtained from the more rigorous
NGF theory with simpler models shows that the spectra tend to
be offset to higher values in the simpler theories.

\end{abstract}


\maketitle

% body of paper here

\section{Introduction}

\begin{figure}
\includegraphics[height=8cm,keepaspectratio]{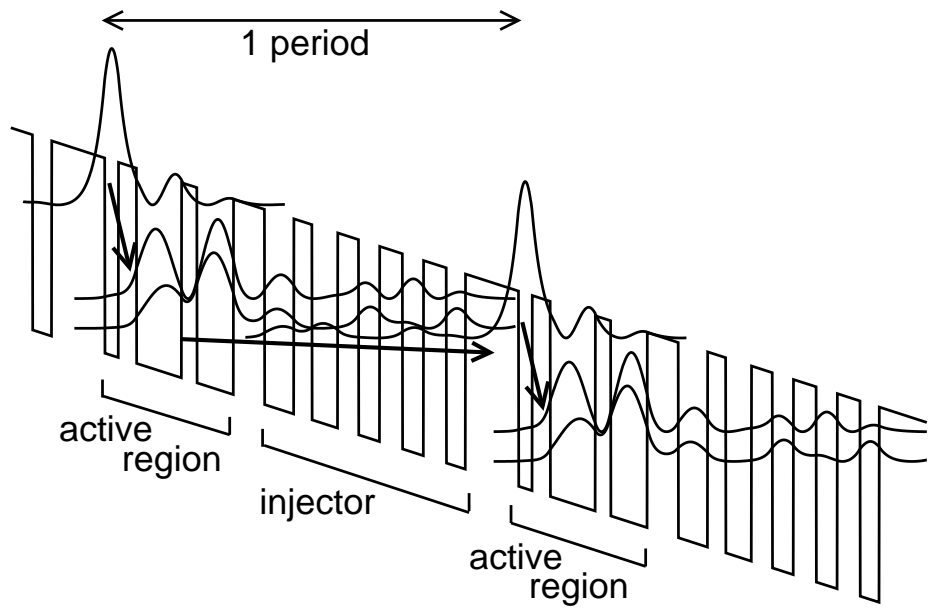}
\caption{Example of the conduction-band lineup in a quantum cascade structure
with an applied bias. The arrows indicate the direction of electron flow
in the structure.
\label{fig.qcstructure}
}
\end{figure}

Quantum cascade (QC) structures are semiconductor heterostructures
grown with a complicated sequence of alternating layers
of different semiconductor materials and with varying thicknesses.
This sequence of layers is repeated many times, up to tens or
even over a hundred periods.
Figure~\ref{fig.qcstructure} shows an example of the conduction-band 
line-up in  a QC structure.
These structures form the basis of a new type of semiconductor
laser,\cite{Fai94b} %FAI94A
in which the laser light emission occurs
through intersubband or interminiband
transitions (in most cases within the conduction
band) rather than interband transitions.
These lasers have a great variety of designs, and a recent review is
given in Ref.~\onlinecite{Gma01}.
Until recently,\cite{Wan01} all quantum 
cascade laser (QCL) structures were designed so that each
period in the structure contains an active region in which the lasing
transition occurs, and a separate injector region. 
The injector acts as a reservoir
of electrons for injection into the active region of the next stage. It
also acts as a collector of electrons from the preceding active region.
The direction of the electron flow is
indicated in Fig.~\ref{fig.qcstructure}, and the electron flow is seen to
resemble a cascade
as the electrons move from one stage to the next when a bias is applied.
Hence, the electron transport through a QCL structure is a complicated 
interplay between relaxation through light emission and scattering
in the active region, and transmission through tunneling and scattering
 in the injector region.\cite{Iot01}

Initially, most theoretical investigations\cite{Iot01,Tor00,Har99a} %HAR99
of transport in
QC structures have focused on the role played by scattering processes
in determining transport properties and the dynamics of the
electron distributions in these structures. Very recently, however, the first
theoretical investigations of quantum transport in these structures
which have considered or incorporated quantum effects beyond a Boltzmann
equation approach
 have been reported.\cite{Wac01,Iot01a}%WAC01A
A theoretical study of quantum transport may be treated
using the density matrix formalism,\cite{Kuh98}
or with a nonequilibrium Green's function (NGF) 
approach.\cite{Kad62,Hau96,Lak97,Wac02}

In the work reported here, we extend the NGF theory described
in Ref.~\onlinecite{Wac02} to the study of quantum transport 
in QC structures. Very early results from this
investigation have been reported in Refs.~\onlinecite{Wac01} %WAC01A
and \onlinecite{Lee02}. In this paper, we present further
and more detailed results from this study. In addition,
we have extended the work further to the problem of evaluating
the gain or absorption spectra of these structures,\cite{Wac02a}
and we also report and discuss results of this work here.

In the following section, we describe the theoretical formulation
that we use to derive the transport properties, and then
the linear optical response of QC structures. In Sec.~\ref{sec.results},
we apply this theory to example QC structures, and describe the
results obtained. The last section contains a summary and conclusion. 

\section{Theoretical Formulation}
\label{sec.theory}

\subsection{Basis states and Hamiltonian}
\label{subsec.basis}

We model the QC structure
as a periodic superlattice structure, in which
each period contains $N_s$ semiconductor layers
with varying thicknesses. The Hamiltonian $\hat{H}$
which we use to model this system may be separated
into two parts: $\hat{H} = \hat{H}_o + \hat{H}_{\rm scatt}$.
$\hat{H}_o$ contains the superlattice potential and a
static electric field $\mathcal{E}$ applied in the
growth direction, i.e., $\hat{H}_o = \hat{H}_{\rm SL}
+ \hat{H}_{\mathcal{E}}$. The Hamiltonian $\hat{H}_{\rm scatt}$
describes the scattering processes included in the theory.

The Hamiltonian is expressed in a set of basis states, 
$\Psi_{{\mathbf k},\alpha}({\bf r},z)
 = (e^{i{\bf k}.{\bf r}}/\sqrt{\mathcal{A}})\psi_\alpha(z)$, 
which we assume separable, although
 this is an approximation when the effective mass is position dependent.
In the plane of the semiconductor layers, the basis functions behave
as plane waves. The normalization constant $\mathcal{A}$
is the sample area in this plane.
The position vector ${\bf r}$ and in-plane wave vector
{\bf k} are two-dimensional (2D) vectors.
In the growth direction, here labeled $z$, there are several possible
choices (see Ref. \onlinecite{Wac02} for a discussion)
for the functions $\psi_\alpha(z)$.
Although the physical results obtained from the theory
should be independent of the specific choice
of these functions, there are different advantages or
drawbacks attached to a given choice. For instance, as we
discuss next, one choice may be more suited to expediting
the numerical computation, while another choice may
more easily allow the extraction of physical information
in a form that can be compared with experimental measurements.
Possible choices are (i)
Bloch functions which are eigenstates of the bare superlattice
potential $\hat{H}_{\rm SL}$, and are spatially extended
across the whole structure. 
(ii) Wannier-Stark  
states which are eigenstates of
$\hat{H}_o$, i.e., of the superlattice potential and the applied bias.
These eigenstates are metastable\cite{Glu02} (non-Hermitian) with
complex energies. They are often treated approximately as
stationary states (their metastable nature is neglected),
and this leads to an ambiguity in the definition of these states
depending on how this approximation is made.
(iii) Wannier functions, which should
not be confused with Wannier-Stark functions, may be constructed so that
they are spatially well localized.
In particular, they may be constructed as eigenstates of the
position operator. In symmetric superlattice structures, these
eigenstates are maximally localized.\cite{Koh59} 
The spatial localization of the Wannier states
enables one to setup a picture of transport in which
scattering occurs from one spatial region of the 
structure to another. The Wannier functions do not depend on
bias, and this is computationally advantageous since coupling
matrix elements between Wannier functions may be calculated
only once at zero bias, and the same matrix elements may then be used at
any applied bias. For these reasons, the theory presented here
is formulated in the Wannier function basis
(see Appendix \ref{app.basis}), 
which we derive as eigenstates of the position operator. 
A disadvantage in using
this basis is that the Wannier functions are not energy eigenstates
of the structure, and it is therefore  difficult to obtain
a physical interpretation in the energy picture or domain. For
instance, there is no one-to-one correspondence between an optical
transition in the structure and a transition between a pair of
Wannier functions. This difficulty may be partially circumvented,
however, at a later stage in the calculation, by transforming
the results obtained into the Wannier-Stark basis. We emphasize again
that the basis choice is not in itself an important issue, i.e., the
physical content of the theory should not depend on this choice, 
but a suitable choice of basis can facilitate the 
numerical computation (e.g., Wannier states), or more easily allow 
physical interpretation (e.g., Wannier-Stark states).

Expressing the Hamiltonian in the Wannier basis we obtain
\begin{eqnarray}
\hat{H}_{\rm SL} & = & \sum_{n,\nu} \sum_{\bf k,s}[E_{\nu{\bf k}}
\hat{a}_{n,{\bf k},s}^{\nu\dagger} \hat{a}_{n,{\bf k},s}^{\nu} \nonumber \\
 & & \qquad+  T_{1}^\nu \, ( \hat{a}_{n+1,{\bf k},s}^{\nu\dagger}
 \hat{a}_{n,{\bf k},s}^\nu \,+ \,\hat{a}_{n-1,{\bf k},s}^{\nu\dagger}
 \hat{a}_{n,{\bf k},s}^\nu )
],
\label{eq.hsl}
\end{eqnarray}
where the index $n$ labels a period in the superlattice, and the index
$\nu$ labels a Wannier function $\psi_\nu(z)$  within a period. 
$\hat{a}_{n,{\bf k},s}^{\nu\dagger}$ and 
$\hat{a}_{n,{\bf k},s}^{\nu}$ are creation and
annihilation operators for an electron with  in-plane wave vector
{\bf k}, and spin index $s$, in the $\nu$th Wannier level, in period $n$. 
As stated earlier,
the Wannier functions are not eigenstates of $\hat{H}_{\rm SL}$, and hence
$T_l^\nu$ represents the off-diagonal couplings between Wannier
levels in different periods, and $E_{\nu{\bf k}}$  represents the diagonal
elements of $\hat{H}_{\rm SL}$ in this basis. We keep only terms
in $T_1^\nu$, i.e., we consider only couplings between adjacent periods.
The next-nearest-neighbor couplings $T_2^\nu$ are two or more orders
of magnitude smaller. 
The Hamiltonian $H_\mathcal{E}$, due to the electric field $\mathcal{E}$,
is written as
\begin{eqnarray}
\hat{H}_\mathcal{E} & = & \sum_{n,\nu,\mu} \sum_{{\bf k},s} \{
-e\mathcal{E}R_0^{\mu\nu}\hat{a}_{n,{\bf k},s}^{\mu\dagger}\, 
\hat{a}_{n,{\bf k},s}^\nu
- ne\,\mathcal{E} 
d \,\delta_{\mu\nu}\hat{a}_{n,{\bf k},s}^{\mu\dagger}\, 
\hat{a}_{n,{\bf k},s}^\mu
\nonumber\\
&& \qquad\qquad\qquad\quad   -\;e\mathcal{E}R_1^{\mu\nu} 
[ \,
\hat{a}_{n+1,{\bf k},s}^{\mu\dagger}\, \hat{a}_{n,{\bf k},s}^\nu +
\hat{a}_{n,{\bf k},s}^{\nu\dagger}\, \hat{a}_{n+1,{\bf k},s}^\mu\,
]
\},
\label{eq.he}
\end{eqnarray}
where  $R_l^{\mu\nu} = \int dz\; \psi^*_\mu(z - ld)\, z \,\psi_\nu(z)$.
$d$ is the length of one period and $e < 0$ is the electron charge.

In the scattering Hamiltonian $\hat{H}_{\rm scatt}$, we include
interface roughness, impurity, and electron-phonon scattering processes.
Both acoustic phonons and longitudinal optical (LO) phonons are
considered. The electron--acoustic-phonon scattering
rates are far smaller than those of the other scattering processes
(Table \ref{tab.rates}), but we include these to provide a channel
for small energy transfers between the electrons and lattice.
This opens up a route 
for the carrier-lattice system to move towards
thermal equilibrium, particularly at low bias when the electrons
may have insufficient energy to emit LO-phonons.

In the NGF theory, the scattering processes expressed in
$\hat{H}_{\rm scatt}$ are treated in the form of self-energies
which are described below.
We have also included electron-electron scattering in a mean-field
approximation so that the interaction between electrons appears
as an additional single-particle potential in $\hat{H}_{\rm scatt}$, 
i.e., we replace $\hat{H}_{\rm scatt}$ with 
$\hat{H}^\prime_{\rm scatt} = \hat{H}_{\rm scatt} + \hat{H}_{\rm MF}$.
$\hat{H}_{\rm MF}$ is written as
\begin{equation}
\hat{H}_{\rm MF} \; = \; \sum_{m\mu,n\nu} \sum_{{\bf k},s}
[V_{\rm MF}]_{m\mu,n\nu}\,\hat{a}_{m,{\bf k},s}^{\mu\dagger}\, 
\hat{a}_{n,{\bf k},s}^\nu,
\label{eq.hmf}
\end{equation}
where $[V_{\rm MF}]_{m\mu,n\nu} = \int dz\; \psi^*_{m\mu}(z)\, 
V_{\rm MF}(z) \,\psi_{n\nu}(z)$.
The evaluation of the mean-field potential $V_{\rm MF}(z)$ is described in the
following subsection.

\subsection{Quantum transport equations and self-energies}
\label{subsec.qteself}

To determine the transport properties of a QC
structure under applied bias, we need first to obtain a
description of the nonequilibrium stationary state of the
system. This information is contained in the nonequilibrium
Green's functions: the retarded Green's function 
${\bf G}^{\rm ret}(E)$ and the correlation function 
${\bf G}^<(E)$ (bold type {\bf G}, and ${\mathbf \Sigma}$
below, indicates a matrix in, e.g., the Wannier basis). 
The quantum transport equations, which
these functions obey, can be derived\cite{Hau96,Wac02} from the Hamiltonian
$\hat{H}$ and have the following form.
The Dyson equation for the matrix element
$G^{\rm ret}_{\alpha_1\alpha_2,{\bf k}}(E)$ 
is written as 
\begin{equation}
E\, G^{\rm ret}_{\alpha_1\alpha_2,{\bf k}}(E)  
 - \sum_{\beta}  \Big[(\hat{H}_o + \hat{H}_{\rm MF})_{\alpha_1\beta,{\bf k}} + 
\Sigma^{\rm ret}_{\alpha_1\beta,{\bf k}}(E) \Big]
G^{\rm ret}_{\beta\alpha_2,{\bf k}}(E) = \delta_{\alpha_1\alpha_2},
\label{eq.gret}
\end{equation}
where $\alpha_1$, $\alpha_2$, and $\beta$ are general indices that include
both the period and Wannier level indices, e.g. $\alpha_1 \equiv (m,\mu)$.
The correlation function $G^<_{\alpha_1\alpha_2,{\bf k}}(E)$ is obtained
from the Keldysh relation:
\begin{equation}
G^<_{\alpha_1\alpha_2,{\bf k}}(E) = \sum_{\beta\beta'} 
G^{\rm ret}_{\alpha_1\beta,{\bf k}}(E)\, \Sigma^<_{\beta\beta',{\bf k}}(E)\, 
G^{\rm adv}_{\beta'\alpha_2,{\bf k}}(E),
\label{eq.gless}
\end{equation}
where ${\bf G}^{\rm adv}(E) = [{\bf G}^{\rm ret}(E)]^\dagger$.

The self-energies ${\mathbf \Sigma}^{\rm ret}(E)$ and
${\mathbf\Sigma}^<(E)$ 
originate from the scattering processes
contained in $\hat{H}_{\rm scatt}$ (i.e., excluding $\hat{H}_{\rm MF}$),
and are evaluated within the self-consistent Born approximation.
Assuming that the diagonal parts of the Green's functions and the
self-energies dominate (because the basis states are
spatially localized),
the self-energy for interface roughness or impurity scattering has the form
\begin{equation}
\Sigma_{\alpha_1\alpha_1,{\bf k}}^{<,{\rm rough/imp}}(E) = \sum_{\beta,{\bf k'}}
\langle |V_{\alpha_1 \beta}^{\rm rough/imp}( {\bf k - k'})|^2  \rangle 
\,G_{\beta \beta,{\bf k'}}^<(E). 
\label{eq.sigrough}
\end{equation}
$V_{\alpha_1\beta}^{\rm rough}( {\bf k' - k} )$ and
$V_{\alpha_1 \beta}^{\rm imp}( {\bf k' - k} )$ are  matrix elements
for interface roughness and impurity
scattering. In the former case, the angle brackets denote an average over
all possible distributions of variations in the interface thickness, and
in the latter case these brackets denote an average over all possible
impurity distributions.
The equation for ${\mathbf \Sigma}^{\rm ret, rough/imp}(E)$ is 
obtained by making the replacements: 
${\mathbf \Sigma}^{<,{\rm rough/imp}}(E) \, 
\rightarrow \,{\mathbf \Sigma}^{\rm ret,rough/imp}(E)$, 
and ${\mathbf G}^{<}(E)\, \rightarrow \, {\mathbf G}^{\rm ret}(E)$. 

For optical or acoustic phonon scattering, the self-energies are
\begin{eqnarray}
 \Sigma_{\alpha_1\alpha_1,{\bf k}}^{\rm ret, phon}(E)\; 
 &= &\;\sum_{\beta,{\bf k'}} \, 
 |V_{\alpha_1 \beta}^{\rm phon}({\bf k},{\bf k'})|^2 \, \nonumber \\
& & \; \times\;\biggr\{
[f_B(E_{\rm phon}) + 1] \,
G_{\beta \beta,{\bf k'}}^{\rm ret}( E - E_{\rm phon} ) \nonumber \\
& &\;+\, f_B(E_{\rm phon})\;
G_{\beta \beta,{\bf k'}}^{\rm ret}( E + E_{\rm phon} ) \nonumber \\
 & &\;+\, 
  \frac{1}{2}\, \Big[ G_{\beta \beta,{\bf k'}}^<( E - E_{\rm phon} ) 
-  G_{\beta\beta,{\bf k'}}^<( E + E_{\rm phon} ) \Big]\nonumber \\
& & +\; i \int \frac{dE_1}{2\pi}\, G^<_{\beta\beta,{\bf k'}}(E -E_1)
\Big( \mathcal{P}\Big\{\frac{1}{E_1 - E_{\rm phon}}\Big\}
- \mathcal{P}\Big\{\frac{1}{E_1 + E_{\rm phon}}\Big\} \Big)
 \biggl\},
\label{eq.sigret}
\end{eqnarray} 
 and
\begin{eqnarray}
&& \Sigma_{\alpha_1 \alpha_1,{\bf k}}^{<, {\rm phon}}(E) = 
\sum_{\beta,{\bf k'}} \, 
|V_{\alpha_1 \beta}^{\rm phon}({\bf k},{\bf k'})|^2 \, \nonumber \\
&& \qquad\qquad\qquad\qquad\quad\times\;\biggr\{ f_B(E_{\rm phon})\, 
G_{\beta \beta,{\bf k'}}^<( E - E_{\rm phon} ) \nonumber \\
&& \qquad\qquad\qquad\qquad\quad + \, [f_B(E_{\rm phon}) + 1]\, 
G_{\beta \beta,{\bf k'}}^<( E + E_{\rm phon} )
\biggl\},
\label{eq.sigless}
\end{eqnarray}
where $V^{\rm phon}$ represents the interaction with optical or 
acoustic phonons, and
$E_{\rm phon}$ represents the energy of the optical or acoustic phonon.
 $f_B(E) = 1/[\exp(E/k_B T) - 1]$ is the equilibrium
phonon distribution at a temperature $T$. This is the only place where
the lattice temperature appears in this theory.
Further detail about the evaluation of the scattering
matrix elements and the self-energies is given in 
Appendix \ref{app.selfe}.
For simplicity in the numerical evaluation,
we neglect the last line containing the principle
value terms in Eq.~(\ref{eq.sigret}).
To further expedite the numerical computation, 
we use momentum-independent scattering matrix
elements $V_{\alpha\beta}^{\rm phon/rough/imp}({\bf k}_{\rm typ},
{\bf k'}_{\rm typ})$, employing a  representative momentum
${\bf k}_{\rm typ}$ (see Appendix \ref{app.selfe}).

As mentioned in the previous section, we include also electron-electron
scattering in the form of a mean-field potential which we determine
by solving Poisson's equation
\begin{equation}
\frac{d^2 V_{\rm MF}(z)}{dz^2} \;=\; 
- \,\frac{e}{\epsilon_s}[\rho_{\rm dope}(z) + \rho_e(z)],
\label{eq.poiss}
\end{equation}
with periodic boundary conditions $V_{\rm MF}(z + d) = V_{\rm MF}(z)$.
$\epsilon_s$ is the absolute static background permittivity.
$\rho_{\rm dope}(z)$ is the dopant density profile in the structure,
and the electron density profile
\begin{equation}
\rho_e(z) \; = \; 2e \sum_{\alpha\beta} \sum_{\bf k}
\int \frac{dE}{2\pi}\,
 -iG^<_{\alpha\beta,{\bf k}}(E)\,\psi^*_\beta(z)\,\psi_\alpha(z),
 \label{eq.rhoe}
\end{equation}
with the factor of 2 from summation over the spin index.

To determine the Green's functions,
${\bf G}^{\rm ret}(E)$ and ${\bf G}^<(E)$, we loop iteratively between
the quantum transport equations,
Eqs.~(\ref{eq.gret}) and (\ref{eq.gless}), and the
equations for the self-energies and mean-field potential
Eqs.~(\ref{eq.sigrough}) -- (\ref{eq.rhoe}),
until we reach a self-consistent solution for these equations.
To solve these equations, we assume that the system is an infinite periodic
structure with period $d$. The boundary conditions $G_{\alpha\beta}(E) =
G_{\alpha'\beta'}(E + e\mathcal{E}d)$ are applied between each period 
to model the bias drop per period in the structure, where $\alpha'$ and $\beta'$
are shifted one period to the left of $\alpha$ and $\beta$.
As starting conditions for the iterative loop, we assume the electrons
are in a Fermi distribution with temperature $T$, and that the density
of states in each Wannier subband is a simple stepfunction.
However, once a converged solution was obtained for a given applied
voltage, we used this solution as a starting point for the calculation
at the next bias to facilitate the convergence process. To test for
convergence, we compared the differences in the ${\bf k}$-integrated
Green's functions $[{\bf G}^{{\rm ret},<}_{\alpha\alpha}(E)]_{i+1} -
[{\bf G}^{{\rm ret},<}_{\alpha\alpha}(E)]_{i}$, and carrier density 
 $[n_e]_{i+1} - [n_e]_i$, evaluated in two successive iteration steps,
with a given tolerance value. A further test for convergence could be
carried out by starting the calculation at different bias points, e.g.,
starting at zero bias and increasing the voltage to a high bias point,
then restarting the calculation at a high bias point and decreasing
the voltage to a low value.

\subsection{Transport properties}

Having determined ${\bf G}^{\rm ret}(E)$ and ${\bf G}^<(E)$, 
information about the stationary state of the system,
such as excitation energies, energy renormalizations,
broadenings or lifetimes, density of states,  populations,
and distribution functions can be extracted. Examples of such information
and their application are given in Sec.~\ref{subsec.neqstate}.
We can also evaluate the current density
using the rate of change of the position operator $\hat{z}$
\begin{equation}
J(t) \; = \; \frac{e}{\mathcal{V}}\langle \frac{d\hat{z}}{dt} \rangle
\; = \; \frac{e}{\mathcal{V}} \frac{i}{\hbar} \langle [\hat{H},\hat{z}] \rangle.
\label{eq.j}
\end{equation}
$J(t)$ is the average current density flowing in the $z$ direction,
 in system volume $\mathcal{V}$.
Recalling the separation of the Hamiltonian $\hat{H}$ into two parts,
$\hat{H}_o$ and $\hat{H}^\prime_{\rm scatt}$, we can determine two separate
contributions to the current $J(t)$. The first contribution comes from
$\hat{H}_o$,
\begin{equation}
J_o(t) \; = \; \frac{e}{\mathcal{V}}\frac{i}{\hbar} 
                  \langle [\hat{H}_o,\hat{z}] \rangle
             \; = \; \frac{e}{\hbar\mathcal{V}} 
	     {\rm Tr}\{[\hat{H}_o,\hat{z}]\, {\bf G}^<(t,t)\},
\label{eq.jot}
\end{equation}
which we evaluate in the energy representation
\begin{equation}
J_o \; = \;  \frac{2e}{\hbar\mathcal{V}}
	     \sum_{\alpha\beta}\sum_{\bf k}\int\frac{dE}{2\pi}\,
	     [\hat{H}_o,\hat{z}]_{\alpha\beta}\,
	     G_{\beta\alpha,{\bf k}}^<(E).
\label{eq.jo}
\end{equation}

\noindent The second contribution is from $\hat{H}^\prime_{\rm scatt}$,
\begin{eqnarray}
J^\prime_{\rm scatt}(t) &=& J_{\rm scatt}(t) + J_{\rm MF}(t) \nonumber \\
            & = & \frac{e}{\mathcal{V}}\frac{i}{\hbar} 
                  \langle [\hat{H}_{\rm scatt}  +
		        \hat{H}_{\rm MF},\hat{z}]\rangle, 
\label{eq.jscattp}			
\end{eqnarray}
where we evaluate $J_{\rm MF}$ similarly to $J_o$, i.e., 
Ref. \onlinecite{foot.jmf}
\begin{equation}
J_{\rm MF} \; = \;  \frac{2e}{\hbar\mathcal{V}}
	     \sum_{\alpha\beta}\sum_{\bf k}\int\frac{dE}{2\pi}\,
	     [\hat{H}_{\rm MF},\hat{z}]_{\alpha\beta}\,
	     G_{\beta\alpha,{\bf k}}^<(E).
\label{eq.jmf}
\end{equation}
The current contribution driven by $H_{\rm scatt}$ 
is written in the energy representation as (see Appendix \ref{app.jscatt})
\begin{eqnarray}
J_{\rm scatt} & = & \frac{2e}{\hbar\mathcal{V}}\sum_{\alpha\beta\gamma}
                 \sum_{\bf k}
	      \int \frac{dE}{2\pi} [ G^<_{\beta\gamma,{\bf k}}(E)
	      \Sigma^{{\rm adv}(\alpha)}_{\gamma\gamma,{\bf k}}(E) +
	      G^{\rm ret}_{\beta\gamma,{\bf k}}(E) 
	      \Sigma^{<(\alpha)}_{\gamma\gamma,{\bf k}}(E)
	      ] z_{\gamma\beta} \nonumber\\
	      &   & \quad -\; z_{\alpha\gamma} [
	      \Sigma^{<(\beta)}_{\gamma\gamma,{\bf k}}(E) 
	      G^{\rm adv}_{\gamma\alpha,{\bf k}}(E)
	      \; + \; 
	      \Sigma^{{\rm ret}(\beta)}_{\gamma\gamma,{\bf k}}(E)
	      G^<_{\gamma\alpha,{\bf k}}(E)
	      ].
\label{eq.jscatt}
\end{eqnarray}
The superscript notation $(\alpha)$ in, for example, the self-energy 
$\Sigma^{<(\alpha)}_{\gamma\gamma,{\bf k}}(E)$ indicates that we take only the
part of the self-energy which depends on $G_{\alpha\alpha}(E)$.
The factor of 2 in Eqs.~(\ref{eq.jo}), (\ref{eq.jmf}), and
(\ref{eq.jscatt}) is from the spin index summation.

  We note here, that although we have divided the current into
separate contributions driven by $\hat{H}_o$, $\hat{H}_{\rm MF}$, and
$\hat{H}_{\rm scatt}$, this should not be taken to mean that
the scattering processes play no role in driving, for example,
the current $J_o$. Although, $\hat{H}_{\rm scatt}$ does not appear explicitly
in Eq.~(\ref{eq.jot}), it is instrumental in determining ${\bf G}^<$, and,
hence, implicitly influences the current $J_o$. 
The scattering processes provide channels for energy dissipation from the
electronic system, so that as the electrons descend downwards through
the electric potential their kinetic energy does not increase but remains
constant, and a steady current flow is maintained through the structure.

\subsection{Gain and absorption spectra}
\label{subsec.gaintheory}

We can also use the information contained in ${\bf G}^{\rm ret}(E)$ and
${\bf G}^<(E)$ as a starting point to evaluate the gain or absorption
spectra of the structure. To do this, we consider the linear response of the
stationary state described by  ${\bf G}^{\rm ret}(E)$ and
${\bf G}^<(E)$ to a small perturbation from an optical field. 
From this response
we determine the susceptibility $\chi(\omega)$, and from Maxwell's
equations\cite{Hau93} we obtain the gain  coefficient
\begin{equation}
g(\omega) \;\simeq\; -\frac{\omega}{c} \frac{{\rm Im}[\chi(\omega)]}{n_B},
\label{eq.gaincof}
\end{equation}
where $n_B \sim \sqrt{\epsilon_s}$ is the background refractive index. 
The applied optical field 
\begin{equation}
\bm{E}({\bf r},t) \; =\; {\bf e}_z
\int\frac{d\omega}{2\pi}\mathcal{E}({\omega})
e^{ik(\omega)y - i\omega t},
\end{equation}
which propagates in the $y$ direction
is included as a small time-dependent
perturbation to the
Hamiltonian: $ \hat{H} \rightarrow \hat{H} + \delta \hat{V}(t)$, where
the perturbation 
\begin{equation}
\delta \hat{V}({\bf r},t) \;=\; 
-\frac{e}{2m_e(z)}\big[\hat{\bf p}\cdot{\bf A}({\bf r},t)\,+\,
{\bf A}({\bf r},t)\cdot\hat{\bf p}\big] \;+\; e\phi({\bf r},t),
\label{eq.vrt}
\end{equation}
with the momentum operator $\hat{\bf p}$, and $m_e(z)$ is the 
spatially dependent effective mass.
The vector potential ${\bf A}({\bf r},t)$ and scalar potential $\phi({\bf r},t)$
are related to the optical field through
\begin{eqnarray}
{\bf A}({\bf r},t) & = & 
\int\frac{d\omega}{2\pi}
\frac{\mathcal{E}({\omega})}{i\omega}e^{ik(\omega)y - i\omega t}\,
{\bf e}_z, \nonumber \\
\phi({\bf r},t) & = & 0,
\label{eq.coulg}
\end{eqnarray}
in the Coulomb gauge.
A discussion of the use of different gauges in this problem is given in 
Ref.~\onlinecite{Wac02b}. For the results presented here, we
apply the long-wavelength approximation,
in which we neglect the spatial variation of the optical field across
the structure. Hence, in this gauge, the perturbation is written
in the energy representation as
\begin{equation}
\delta{V}_{\alpha\beta}(\omega) \; = \; 
i \frac{e\mathcal{E}(\omega)}{\omega}
\bigg[\frac{\hat{p}_z}{m_e(z)}\bigg]_{\alpha\beta}
\; = \; - \frac{e\mathcal{E}(\omega)}{\hbar\omega}
[\hat{H}_o,\hat{z}]_{\alpha\beta}.
\end{equation}

The linear changes in the Green's functions and self-energies
caused by the additional term $\delta \hat{V}(t)$ in the Hamiltonian
represent the linear
response of the nonequilibrium stationary state to the applied optical field.
These changes may be written as\cite{Wac02b}
\begin{equation}
\delta {\bf G}^{\rm ret}(\omega,E)  =  
{\bf G}^{\rm ret}(E +\hbar\omega)[\delta {\bf V}(\omega)
+ \delta {\mathbf \Sigma}^{\rm ret}(\omega,E)]{\bf G}^{\rm ret}(E) 
\label{eq.delgret}
\end{equation}
\begin{equation}
\delta {\bf G}^{\rm adv}(\omega,E)  = 
{\bf G}^{\rm adv}(E +\hbar\omega)[\delta {\bf V}(\omega)
+ \delta {\mathbf \Sigma}^{\rm adv}(\omega,E)]{\bf G}^{\rm adv}(E), 
\end{equation}
and
\begin{eqnarray}
\delta {\bf G}^<(\omega,E) & = & 
                       {\bf G}^{\rm ret}(E +\hbar\omega)\,
		       \delta {\bf V}(\omega)\,
                      {\bf G}^<(E) \, +\, {\bf G}^<(E+\hbar\omega)
		      \,\delta {\bf V}(\omega)\,
		    {\bf G}^{\rm adv}(E) \nonumber \\
	& & + \;{\bf G}^{\rm ret}(E +\hbar\omega)\,
	          \delta{\mathbf \Sigma}^{\rm ret}(\omega,E)\,
	     {\bf G}^<(E) + {\bf G}^{\rm ret}(E +\hbar\omega)\,
	     \delta{\mathbf \Sigma}^<(\omega,E)\,
	     {\bf G}^{\rm adv}(E) \nonumber \\
	& & +\; {\bf G}^<(E +\hbar\omega)\,
	         \delta{\mathbf \Sigma}^{\rm adv}(\omega,E)\,
	     {\bf G}^{\rm adv}(E).
\label{eq.delgless}
\end{eqnarray}
Note that $\delta {\bf G}^{\rm adv}(\omega,E) 
\neq [\delta {\bf G}^{\rm ret}(\omega,E)]^\dagger$.
The expressions for $\delta{\mathbf \Sigma}(E)$ 
take the same form as in Eqs.~(\ref{eq.sigrough})
-- (\ref{eq.sigless}) where the functions ${\bf G}(E)$ are replaced by 
$\delta {\bf G}(E)$.
For $\delta{\mathbf \Sigma}^{\rm adv}(E)$, 
the expression for $\delta{\mathbf \Sigma}^{\rm ret}(E)$
is used with the replacement $\delta {\bf G}^<(E) 
\rightarrow -\delta {\bf G}^<(E)$ and
$\delta {\bf G}^{\rm ret}(E) \rightarrow \delta {\bf G}^{\rm adv}(E)$.
$\delta {\mathbf \Sigma}^{</{\rm ret}}(E)$ 
must be evaluated self-consistently with
$\delta {\bf G}^{</{\rm ret}}(E)$, 
and therefore another iterative loop must be carried
out, similar to the earlier
procedure to determine ${\bf G}^{</{\rm ret}}(E)$ and 
${\mathbf \Sigma}^{</{\rm ret}}(E) $.
From the change in the Green's functions and self-energies,
 and using Eqs.~(\ref{eq.j}) -- (\ref{eq.jscatt}), we
obtain the change in current density
\begin{equation}
\delta J(\omega) = \delta J_o(\omega) +\delta J_{\rm MF}(\omega) +
                              \delta J_{\rm scatt}(\omega),
\label{eq.deltaj}
\end{equation}
where\cite{foot.jmf}
\begin{eqnarray}
\delta J_o(\omega) & = & \frac{2e}{\hbar\mathcal{V}}
                           \sum_{\alpha\beta}\sum_{\bf k}
                        \int \frac{dE}{2\pi}
                       [\hat{H}_o,\hat{z}]_{\alpha\beta}\; 
		      \delta {G}_{\beta\alpha,{\bf k}}^<(\omega,E) 
                         + [\delta {V}, \hat{z}]_{\alpha\beta}\; 
			 {G}_{\beta\alpha,{\bf k}}^<(\omega, E),
\label{eq.deljo}
\end{eqnarray}
\begin{eqnarray}
\delta J_{\rm MF}(\omega) & = & \frac{2e}{\hbar\mathcal{V}}
                         \sum_{\alpha\beta}\sum_{\bf k} 
                     \int \frac{dE}{2\pi}
                      [\hat{H}_{\rm MF},\hat{z}]_{\alpha\beta}\; 
		      \delta {G}_{\beta\alpha,{\bf k}}^<(\omega,E),
\label{eq.deljmf}
\end{eqnarray}
and
\begin{eqnarray}
\delta J_{\rm scatt}(\omega) & = & \frac{2e}{\hbar\mathcal{V}} 
                         \int \frac{dE}{2\pi}
                      \sum_{\alpha\beta\gamma}\sum_{\bf k}\,
	       \biggl[ \delta G^<_{\beta\gamma,{\bf k}}(\omega,E)
	      \Sigma^{{\rm adv}(\alpha)}_{\gamma\gamma,{\bf k}}(E) +
	      \delta G^{\rm ret}_{\beta\gamma,{\bf k}}(\omega,E) 
	      \Sigma^{<(\alpha)}_{\gamma\gamma,{\bf k}}(E) \nonumber \\
	  & &  \qquad\qquad  +\; G^<_{\beta\gamma,{\bf k}}(E + \hbar\omega)\,
	      \delta\Sigma^{{\rm adv}(\alpha)}_{\gamma\gamma,{\bf k}}(\omega,E) 
	      +	      G^{\rm ret}_{\beta\gamma,{\bf k}}(E+\hbar\omega) \,
	      \delta\Sigma^{<(\alpha)}_{\gamma\gamma,{\bf k}}(\omega,E)
	      \biggr] z_{\gamma\beta} \nonumber\\
	  &   & \quad -\;\sum_{\alpha\beta\gamma}\sum_{\bf k}
	                z_{\alpha\gamma} \biggl[
	      \delta\Sigma^{<(\beta)}_{\gamma\gamma,{\bf k}}(\omega,E) 
	        G^{\rm adv}_{\gamma\alpha,{\bf k}}(E)
	      \; + \; 
	      \delta\Sigma^{{\rm ret}(\beta)}_{\gamma\gamma,{\bf k}}(\omega,E)
	      G^<_{\gamma\alpha,{\bf k}}(E)
	      \nonumber \\
	 & & \qquad\quad +
	 \;\Sigma^{<(\beta)}_{\gamma\gamma,{\bf k}}(E+\hbar\omega) \,
	           \delta G^{\rm adv}_{\gamma\alpha,{\bf k}}(\omega,E)
	      \; + \; 
	      \Sigma^{{\rm ret}(\beta)}_{\gamma\gamma,{\bf k}}(E+\hbar\omega)\,
	      \delta G^<_{\gamma\alpha,{\bf k}}(\omega,E)
	      \biggr],
\label{eq.deljscatt}
\end{eqnarray}
where the factor 2 is again from  the spin index summation.
From the change in current density $\delta J(\omega)$, we obtain the
polarization $\delta P(\omega) = i\delta J(\omega)/\omega$ induced
in the material, and hence the complex susceptibility
\begin{equation}
\chi(\omega) \; = \; \frac{\delta P(\omega)}{\epsilon_0 \mathcal{E}(\omega)}
\; = \; \frac{i}{\epsilon_0}\frac{\delta J(\omega)}{\omega \mathcal{E}(\omega)},
\label{eq.sus}
\end{equation}
and from Eq.~(\ref{eq.gaincof}) we obtain the gain coefficient $g(\omega)$.

\section{Application to quantum cascade structures}
\label{sec.results}

We have applied the theory presented above to some example QCL
structures reported in the literature. These structures
were grown in the GaAs/Al$_x$Ga$_{1-x}$As material system,
but this theory is also applicable to other material systems.
The examples we consider are
[A] the first GaAs/Al$_x$Ga$_{1-x}$As
QCL structure\cite{Sir98} reported by Sirtori {\it et al.},
with $x = 0.33$, and
[B] a structure\cite{Pag01} reported by Page {\it et al.}
    with a similar layer design to structure A but
with a higher Al content in the barrier, $x=0.45$, resulting in
higher barriers.  This QCL structure operates in pulsed mode
at room temperature. The lasing transition in  structures A and B
occurs between quantum well subbands. Other parameters for these
structures are summarized in Table~\ref{tab.samplepar}.
In this section, we present and discuss results obtained
for these structures. 

\subsection{Nonequilibrium stationary state}

\label{subsec.neqstate}

\begin{figure}
\includegraphics[height=12cm,keepaspectratio]{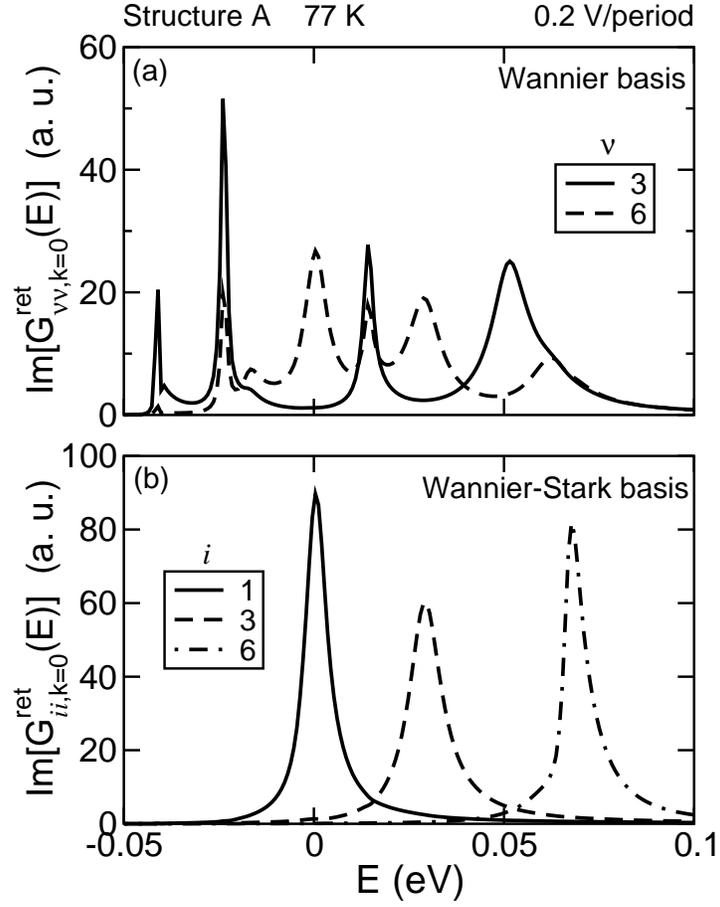}
\caption{Examples of the diagonal elements 
${\rm Im}[G^{\rm ret}_{\nu\nu,{\bf k} = 0}(E)]$ in one period of
structure A with an applied voltage of 0.2 V/period. a) Wannier
basis. b) Wannier-Stark basis.
\label{fig.gretek}
}
\end{figure}

In this subsection, we describe some of the information about the
nonequilibrium stationary state of a QC structure under an applied
bias that can be extracted from the Green's functions. In Fig.~\ref{fig.gretek},
we show some examples of the diagonal elements of
${\rm Im}[{\bf G}^{\rm ret}(E)]$ for ${\bf k} = 0$ plotted as a function of the
energy parameter $E$. 
Figure~\ref{fig.gretek}(a) shows two examples of
this function evaluated in the Wannier basis. 
These examples were evaluated
for structure A at a bias of 0.2 V/period.
Each curve represents a Wannier state in one period of the QC structure. 
The Wannier states are not energy eigenstates of the system, and if a Wannier
state is expressed as a linear superposition of the energy eigenstates,
the position of the peaks in, for instance, the curve labeled $\nu = 3$
represent the energies of the eigenstates which comprise this superposition.

It is possible also to express the Green's functions in the Wannier-Stark
basis. From the diagonalization of the Hamiltonian $\hat{H}_o$ expressed
in the Wannier basis we obtain the  transformation matrix between the
Wannier and Wannier-Stark basis.\cite{foot.ws}
Applying this transformation matrix
to ${\bf G}^{\rm ret}(E)$ we obtain the example curves shown in 
Fig.~\ref{fig.gretek}(b). Each of these curves (diagonal elements of
${\rm Im}[{\bf G}^{\rm ret}(E)]$ for ${\bf k} = 0$ in the Wannier-Stark basis)
represents a Wannier-Stark state.
In contrast to the curves shown
in Fig.~\ref{fig.gretek}(a) with their complicated structure of peaks, 
each curve in the Wannier-Stark basis consists of a  single, broadened
peak. This indicates that at this bias, the Wannier-Stark states
approximate closely the true energy eigenstates of the system.
The  positions of these peaks $E_i$ are shifted downwards by around
10 to 15 meV in comparison to the eigenenergies obtained from
the diagonalization of $\hat{H}_o$. This shift represents
the renormalization of the Wannier-Stark energy levels due to
the scattering processes described in $\hat{H}^\prime_{\rm scatt}$. The
broadening of the peaks also originates from these scattering
processes, and the width of the peaks gives a measure of the
decay rate $\Gamma_i$ of these levels due to scattering. In particular,
we take $\Gamma_i = \Delta E_{\rm fwhm}$, where $\Delta E_{\rm fwhm}$
is the full width at half  maximum of the single peak in the function
${\rm Im}[G^{\rm ret}_{ii,{\bf k}=0}(E)]$, in the Wannier-Stark basis.
In Sec.~\ref{subsec.gainresult}, we will use this information to
analyze the gain spectra obtained from the NGF theory.

\begin{figure}
\includegraphics[height=12cm,keepaspectratio]{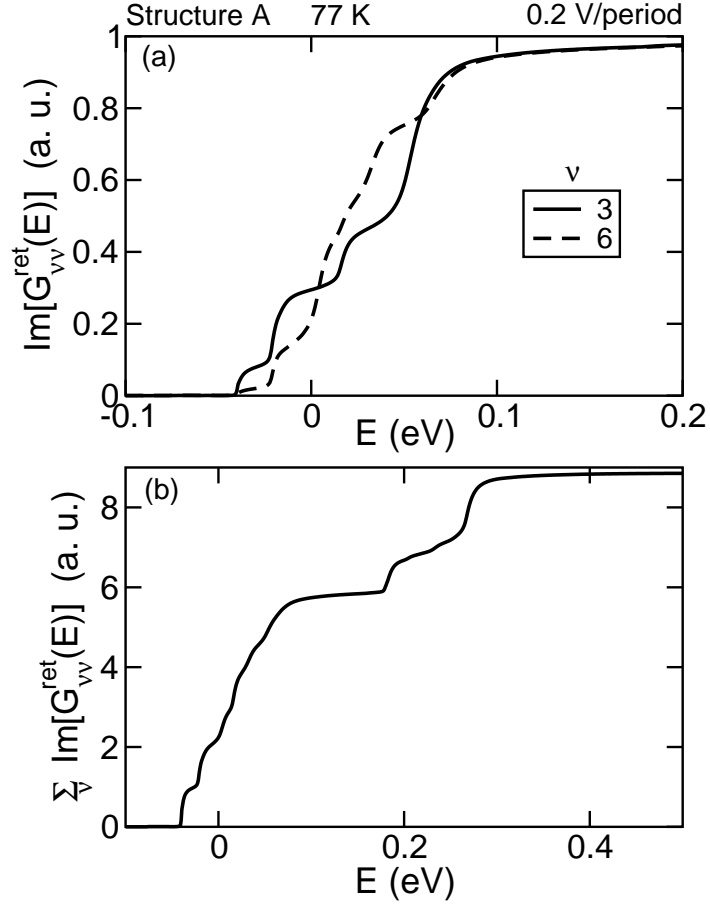}
\caption{ a) Examples of density of states (per period)
in  Wannier subbands: 
${\rm Im}[G^{\rm ret}_{\nu\nu}(E)]
= 2\sum_{\bf k}\; {\rm Im}[G^{\rm ret}_{\nu\nu,{\bf k}}(E)]$.
b) Total density of states per period,
$\sum_\nu {\rm Im}[G^{\rm ret}_{\nu\nu}(E)]$.
\label{fig.dos}
}
\end{figure}

The curves in Fig.~\ref{fig.gretek} were evaluated for ${\bf k} = 0$. If
instead we integrate over the in-plane wave vector ${\bf k}$, we obtain
the curves shown in Fig.~\ref{fig.dos}(a).  
These staircaselike curves are the density of states (DOS)
in two example Wannier subbands (the two Wannier
states considered in Fig.~\ref{fig.gretek}(a) are the ${\bf k} = 0$ states
in these subbands). The onset of each step in the staircase corresponds
to the position of each peak in the corresponding curve
in Fig.~\ref{fig.gretek}(a).
If we carried out the  ${\bf k}$ integral in the Wannier-Stark basis, we would
obtain a single step instead of a staircaselike structure, corresponding
to the single peak (for a given curve) seen in Fig.~\ref{fig.gretek}(b).
If we sum the DOS curves for all Wannier subbands within
a period, we obtain Fig.~\ref{fig.dos}(b), which is the total DOS
per period of the structure. The result in Fig.~\ref{fig.dos}(b)
is basis independent if we sum over contributions from all basis
states within the energy range shown.

\begin{figure}
\includegraphics[height=12cm,keepaspectratio]{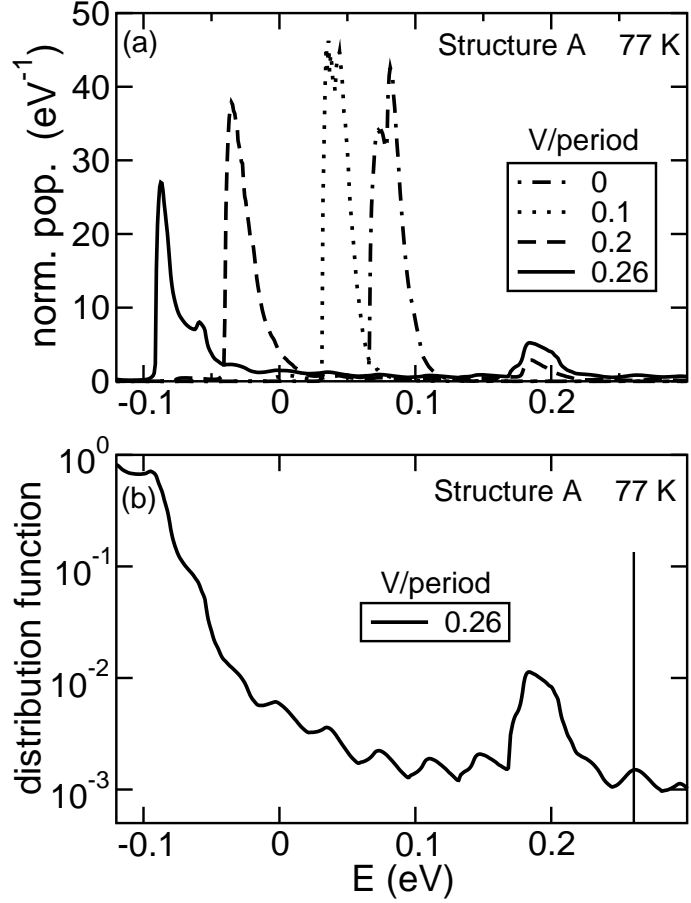}
\caption{ a) $n(E)$, normalized population per period 
for different applied voltages.
b) $f(E)$, distribution function in one period at 0.26 V/period.
The thin vertical line marks the position of the conduction-band
offset between the thin (first) well in the active region in this
period, and the thick barrier separating this well from the preceding
period. The bottom of this well is at around 0 eV.
The peak at around 0.2 eV corresponds to a population inversion
in the upper laser level in the active region.
\label{fig.pop}
}
\end{figure}

We consider now the lesser correlation function ${\bf G}^<(E)$.
At equal times,  $G^<_{\nu\nu,{\bf k}}(t)
= i\langle \hat{a}^\dagger_{\nu{\bf k}}(t) \hat{a}_{\nu{\bf k}}(t)\rangle =
 i \langle \hat{n}_{\nu{\bf k}}(t)\rangle$, where 
 $\langle \hat{n}_{\nu{\bf k}}(t)\rangle$
is the occupation of a state {\bf k} in subband $\nu$. 
If we move into the energy representation, and sum over the contributions 
from all states in all subbands in a period, we obtain the 
population $n(E) = 2\sum_{\nu {\bf k}}
\, G^<_{\nu\nu,{\bf k}}(E)$ as a function of energy within the period.
Figure~\ref{fig.pop}(a) shows the normalised population per period per unit
energy for several applied voltages. 
Dividing these curves by the total DOS curves
for the corresponding bias gives the distribution function
as a function of energy, i.e.,
\begin{equation}
f(E) \; = \; \frac{\sum_{\nu{\bf k}}
\, G^<_{\nu\nu,{\bf k}}(E)}{2\sum_{\nu{\bf k}} 
{\rm Im}[G^{\rm ret}_{\nu\nu,{\bf k}}(E)]}.
\end{equation}
The distribution function
at 0.26 V/period in structure A 
is shown in Fig.~\ref{fig.pop}(b). In this curve, the 
jagged points between 0 and 0.2 eV are LO-phonon replicas, and the onset
of a small population inversion at around 0.2 eV is also seen.

\subsection{Transport}

\begin{figure}

\includegraphics[height=10cm,keepaspectratio]{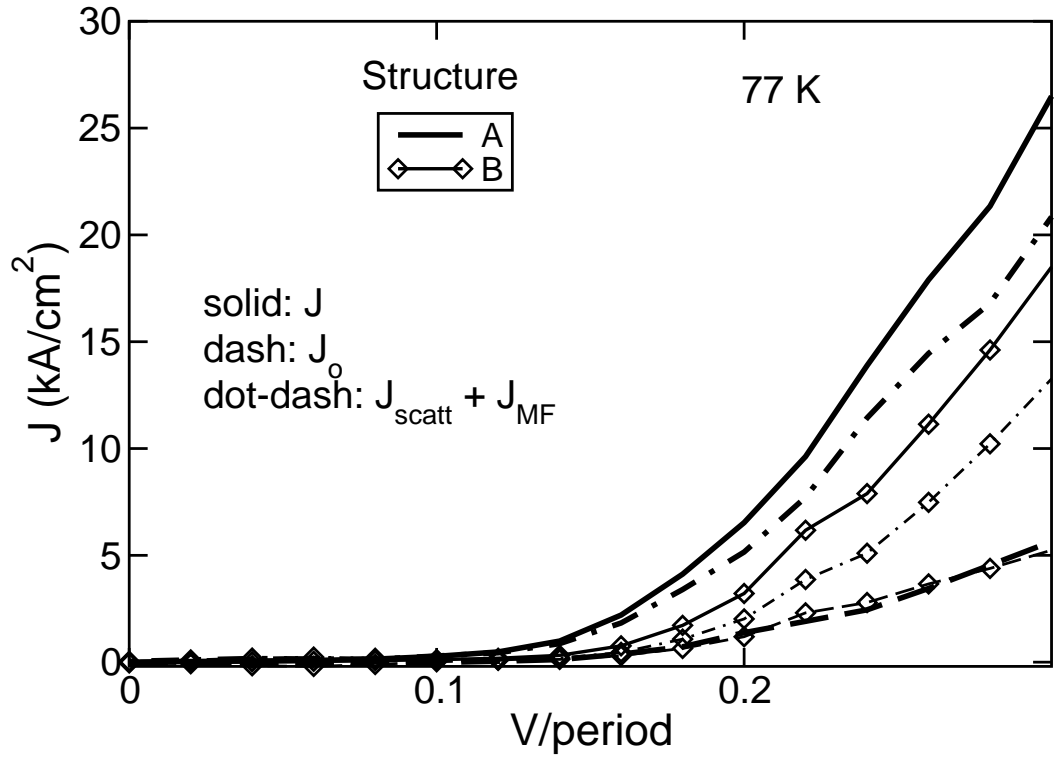}
\caption{Current density vs. voltage/period for example QC structures.
Structure A: Thick lines. Structure B: Thin lines with symbols.
\label{fig.jvab}
}
\end{figure}

We consider now the transport properties of the QC structures
described above.
Figure~\ref{fig.jvab} shows a plot of current density against V/period,
calculated from Eqs.~(\ref{eq.j}) -- (\ref{eq.jscatt}). 
The current density for structure B
 is reduced below  structure A because of the higher barrier
heights in B. 
The separate current contributions $J_o$ and $J_{\rm scatt}
+ J_{\rm MF}$ from
Eqs.~(\ref{eq.jo}) -- (\ref{eq.jscatt}) are also shown. 
In both the structures,
the main contribution is from $J_{\rm scatt}$ ($J_{\rm MF}$ is negligible in 
comparison with $J_{\rm scatt}$). 
This finding is in contrast to the behavior in simple superlattice structures
where the dominant contribution to the current is from $J_o$. This difference
could perhaps be explained by the far greater number of subbands within
one period of a QC structure (in comparison to one period
of a simple symmetric superlattice structure), 
in which the envelope functions $\psi_\alpha(z)$
are spatially displaced across the period. This opens up the 
possibility for many more scattering transitions which facilitates the 
transport of carriers across the structure through scattering processes.
Comparing A and B,
we see that $J_o$ is almost identical for both structures. The
current $J_{\rm scatt}$ in B is around half that in A so that the reduction
in total current in B compared to A arises from a decrease in the transport
driven by the scattering Hamiltonian $\hat{H}_{\rm scatt}$.

\begin{figure}

\includegraphics[height=12cm,keepaspectratio]{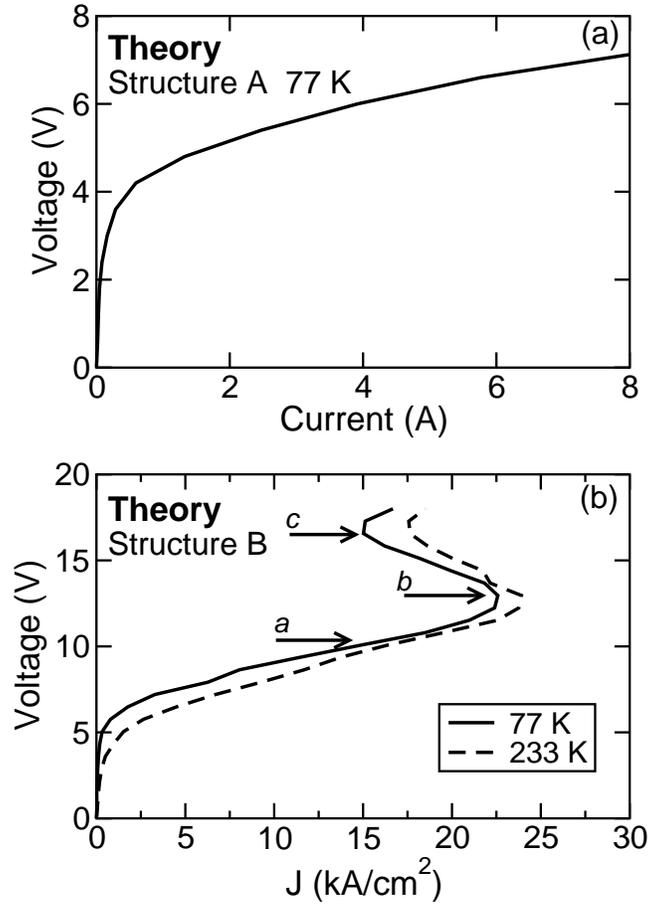}
\caption{Current-voltage characteristic for comparison with experimental data 
for structures A and B (see text for references). 
The alignment of the injector level and upper laser
level at the positions marked with arrows in (b)
is shown in Fig. \ref{fig.wspage}.
\label{fig.viexp}
}
\end{figure}

To compare the theoretical results with experimental data reported in
the literature, we show in Fig.~\ref{fig.viexp}
the results for A and B plotted against similar scales
as the experimental plots shown in Fig.~3 in Ref.~\onlinecite{Sir98},
and Fig.~4 in Ref.~\onlinecite{Pag01}. Comparing the results for
structure A with the data given in Ref.~\onlinecite{Sir98}, we see the 
voltage trend in both curves agree well. In the theory, the voltage
tends to be lower for a given current than in the experiment, but this
may be accounted for by an additional series resistance in the cladding layers.
For structure B, we see again that the voltage trends agree well
with the data shown in Ref.~\onlinecite{Pag01}. In particular, when comparing
the trends at different temperatures, we find that in both the theory
and experiment the current is higher for a given voltage at the higher
temperature. In both theory and experiment, the 
curves at both temperatures converge at a certain current density.
Unlike the experiment, the two curves do not cross at the point of convergence 
in the theory, but diverge beyond this point. The theoretical
curves reproduce the sharp break in the voltage trend at around
23 kA/cm$^2$ as seen in the experiment, and the theory shows that
this break corresponds to the onset of a region of negative differential
resistivity (NDR). NDR is not seen explicitly in the experimental
data in Ref.~\onlinecite{Pag01} because of an artefact of the 
measurement system.\cite{foot.pagendr}

\begin{figure}
\includegraphics[height=12cm,keepaspectratio]{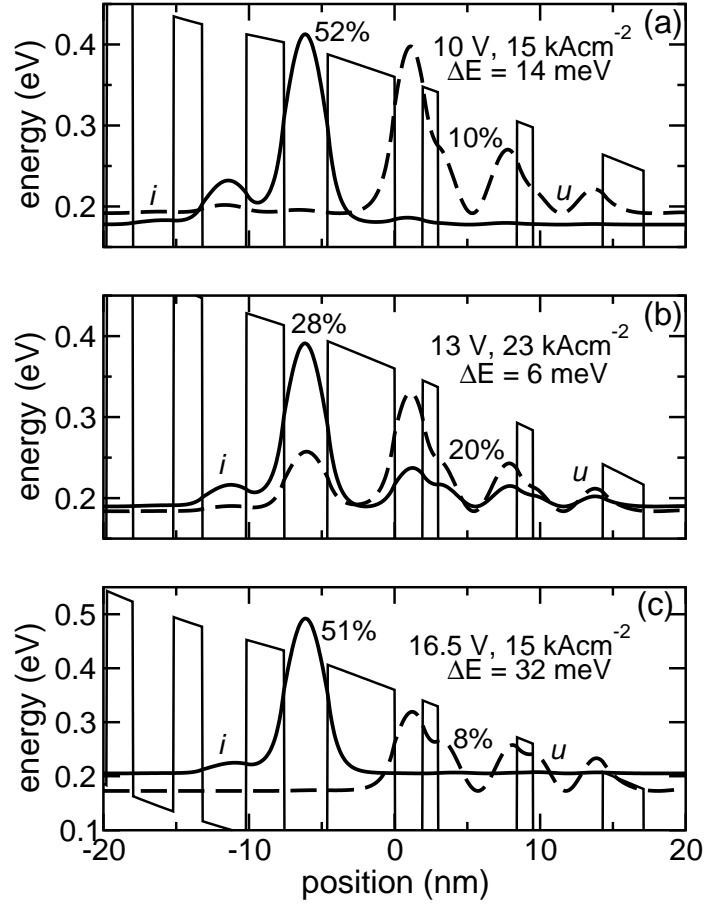}
\caption{ 
Wave functions (mod. squared) of the injector level {\em i} (solid line) 
and upper laser level {\em u} (dashed line) in structure B 
for different applied bias
at 77 K. The energetic positions show the alignment of these levels.
Parts (a) -- (c) correspond to each of the 
voltages marked with arrows in 
Fig.~\ref{fig.viexp}(b). The percentage of population in each level is
also shown. $\Delta E$ is the energy separation between
the two levels.
\label{fig.wspage}
}
\end{figure}

The break in the voltage trend is attributed\cite{Pag01} by Page {\it et al.} 
to the breaking of the level alignment  in the injector region and the 
active region causing an interruption in the current flow. 
To verify this, we show the wave functions and alignment 
of the injector level {\em i}
and upper laser level {\em u} (in the Wannier-Stark basis\cite{foot.ws})
at 77 K,
for different applied voltages, in Fig.~\ref{fig.wspage}.
Figures \ref{fig.wspage}(a) -- (c) correspond to the voltages marked
{\em a}, {\em b}, and {\em c} in Fig.~\ref{fig.viexp}(b). 
Figure \ref{fig.viexp}(b)
shows that as the voltage increases from 10 V, 
the current-voltage characteristic
passes through a peak at 13 V, and then beyond this point the current
decreases to a minimum at around 16.5 V.
Comparing this behavior with Fig.~\ref{fig.wspage}, we see that
the change in voltage, and the behavior of the I-V characteristic,
is accompanied by a shift in the position of
 level {\em i} from below to above level {\em u}. Level {\em i}
passes through the resonance with level {\em u} which causes
the peak in the I-V characteristic seen at 13 V. The population
percentage in these levels is also shown in Fig.~\ref{fig.wspage},
and we see that away from the resonance position, $\sim 50$\%
of the population is in the injector level, with only $\sim 10$\% in the
upper laser level. Close to resonance, however, 
we see that the population is more evenly
distributed between the two levels, with $\sim 30$\% in the injector
level, and 20\% in the upper laser level.  
The wave functions in the near-resonance case (b)
are less well-localized than in the off-resonance cases 
(a) and (c),
and tend to spread across both the injector and active region.
This increases the overlap of the two wave functions and facilitates
the transfer of carriers between the injector and active region.

\begin{figure}
\includegraphics[height=10cm,keepaspectratio]{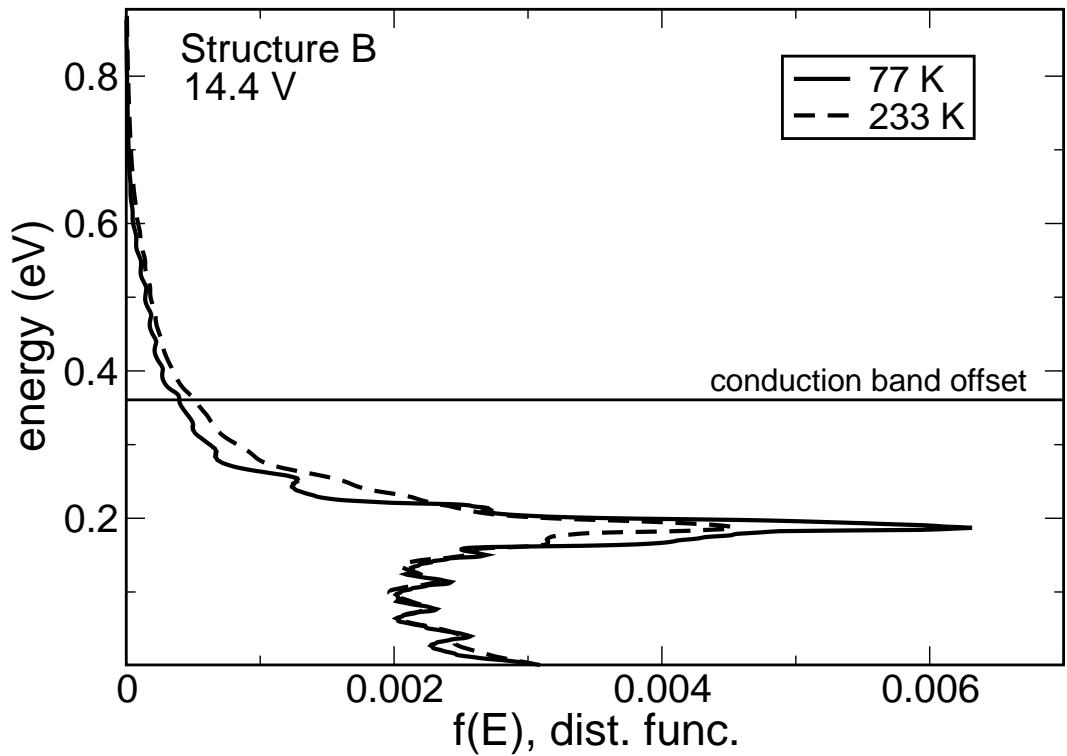}
\caption{ 
Electron distribution functions $f(E)$ at 77 K and 233 K in one period
of structure B at 14.4 V.
\label{fig.distV0.4}
}
\end{figure}

In the experimental data at 233 K, the break in the voltage trend
is less pronounced, and this was attributed\cite{Pag01} to
the presence of electrons with higher energy leaking into continuum states,
and maintaining the current flow. In the 
theoretical results, shown in Fig.~\ref{fig.viexp}(b), there is a small
increase in current at 233 K, but it is difficult to make a quantitative
comparison with the experimental data because of the uncertainty in
the experimental data in the NDR region.\cite{foot.pagendr}
Figure~\ref{fig.distV0.4} shows the electron distribution functions $f(E)$
at 77 K and 233 K, in one period of the structure at 14.4 V, i.e.,
just beyond the peak in the I-V characteristic in Fig.~\ref{fig.viexp}(b).
The thin horizontal line marks the conduction-band offset or barrier height.
The peak in the distributions at around 0.2~eV
corresponds to the population inversion in the upper laser level.
The population inversion at 233 K is about a third less than that
at 77 K. In the high-energy tails, the
distribution is slightly larger at 233 K than at 77 K. 
The difference in the distribution functions is not large, however, 
and although this difference may result in more electron leakage into the
continuum at 233 K, it is not clear at this point if this
is a sufficiently large effect to elucidate the experimental results.

\subsection{Gain and absorption spectra}
\label{subsec.gainresult}

In this section, we apply the theory described in Sec.~\ref{subsec.gaintheory}
to structure A. We distinguish here between
the gain coefficient $g(\omega)$ evaluated using
$\delta J(\omega)$ from Eq.~(\ref{eq.deltaj}), and the gain coefficient
$g_o(\omega)$ which is evaluated neglecting $\delta J_{\rm scatt}(\omega)$
in Eq.~(\ref{eq.deltaj}), and neglecting terms containing 
$\delta {\mathbf \Sigma}^{</{\rm ret}/{\rm adv}}$
in $\delta {\bf G}^<(\omega,E)$ [Eq.~(\ref{eq.delgless})]. 
Hence, $g_o(\omega)$ does not depend on changes in the self-energies, and
it is simpler to evaluate since
its evaluation does not require a further self-consistent calculation.

\begin{figure}

\includegraphics[height=10cm,keepaspectratio]{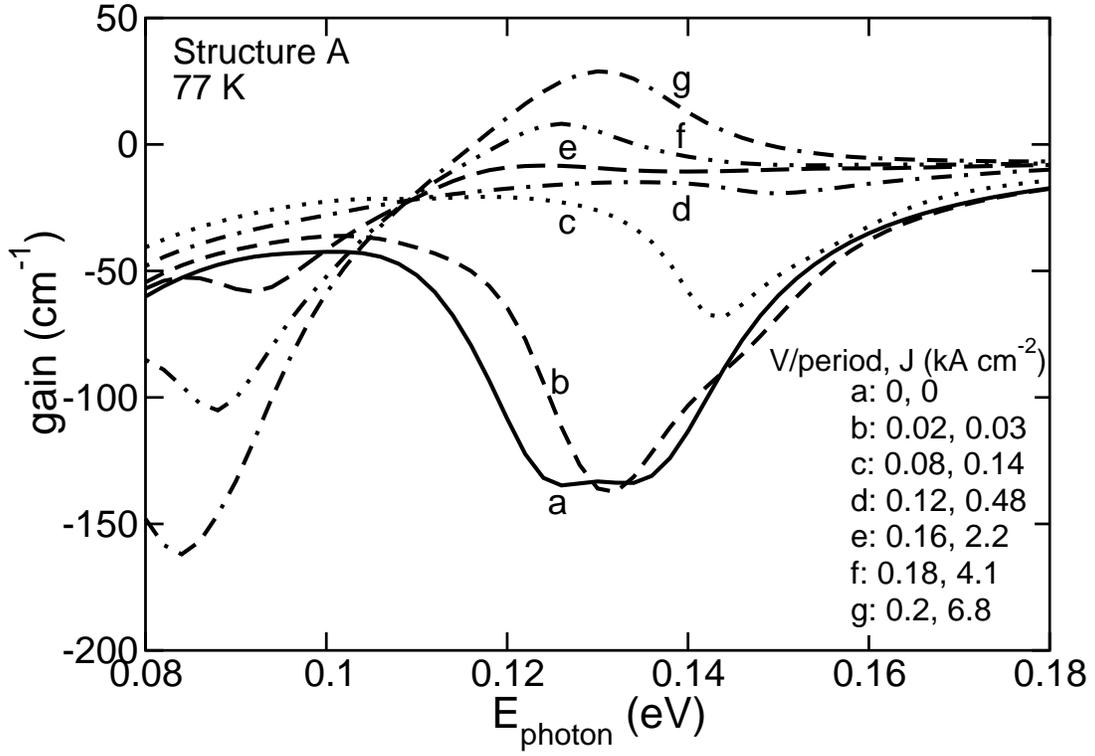}
\caption{Gain and absorption spectra $g(\omega)$ 
for structure A with different applied
voltages, calculated with the NGF theory.
\label{fig.gainNGF}
}
\end{figure}

In Fig.~\ref{fig.gainNGF}, we show the gain coefficient $g(\omega)$ calculated
with the  NGF theory for different applied voltages, between 0
to 0.2 V/period. With zero applied bias, there is a broad absorption
ranging from around 120 to 140 meV. As the applied voltage increases,
the absorption gradually decreases in this range. There is a corresponding
slow increase in absorption between around 80 to 100 meV. At around 0.18
V/period,  a positive gain begins to appear  in the range
120 to 140 meV, and this gain increases further as the voltage is
further increased. These results were obtained in the Wannier
basis and as stated in Sec.~\ref{subsec.basis}, there is a problem
in interpreting the origin of the different features in the 
spectra because these cannot be related to transitions between
Wannier states. Later in this section, we discuss a way round this
problem by considering a transformation to Wannier-Stark states.

\begin{figure}

\includegraphics[height=10cm,keepaspectratio]{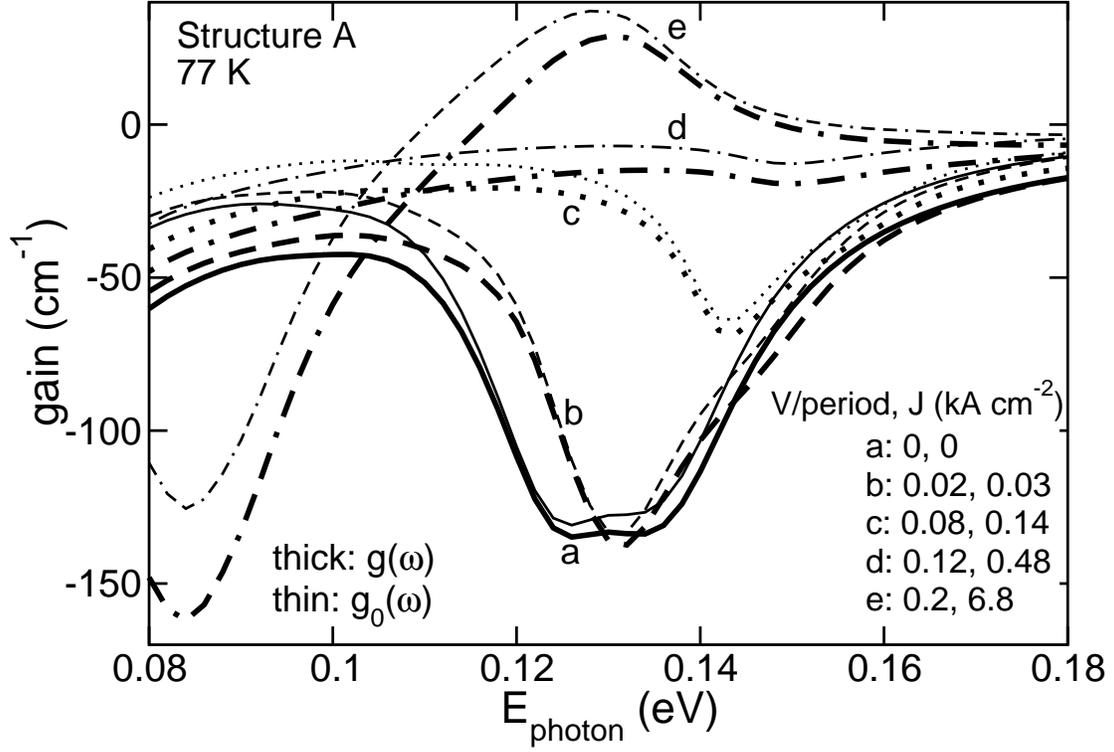}

\caption{Comparison of $g(\omega)$ and $g_o(\omega)$, with different applied
bias, for structure A. a: solid lines, b: dashed lines, c: dotted lines, 
d: dot-dashed lines, e: dot--double-dashed lines.
\label{fig.gainNGFcomp}
}
\end{figure}

Before this we compare the results obtained for $g(\omega)$
from the more rigorous NGF theory, with the gain $g_o(\omega)$ which is
calculated neglecting terms from $\delta J_{\rm scatt}$. This
comparison is shown in Fig.~\ref{fig.gainNGFcomp}. We see that
the simpler theory gives rise to gain curves which are offset to
higher values (i.e., absorption is reduced and gain is increased)
in comparison to the more rigorous theory.

To determine the origin of the different features
in the spectra, we consider a different approach
for evaluating the gain coefficient. In this approach, we 
transform the Green's functions ${\bf G}^{</{\rm ret}}$ obtained
in the Wannier basis to the Wannier-Stark basis 
(${\bf G}^{</{\rm ret}}_{\rm WS}$) as described
earlier in Sec.~\ref{subsec.neqstate}.
 From the diagonal elements
of $G_{{\rm WS},ii}^<$, we obtain the populations $n_i$ 
of the Wannier-Stark levels, and from the positions and 
widths of the single peaks in ${\rm Im}[G_{{\rm WS},ii}^{\rm ret}]$,
we obtain the Wannier-Stark level energies, $E_i$,  for ${\bf k} = 0$,
and the corresponding decay rates, $\Gamma_i$. 
We treat each pair of Wannier-Stark subbands 
in a simple two-band model, in which the optical response due to
subbands $i$ and $j$ is given by,\cite{Hau93}
\begin{equation}
{\rm Im}[\chi_{ij}(\omega)] \; = \; \frac{2\pi}{\epsilon_0\mathcal{V}}
\sum_{\bf k} |d_{ij}|^2\, (f_{i,{\bf k}} - f_{j,{\bf k}})\, \delta(\hbar\omega
+ E_{i,{\bf k}} - E_{j,{\bf k}}),
\end{equation}
where $d_{ij} = e\int dz\,\psi^*_i(z)\, z \,\psi_j(z)$ is the dipole matrix
element between the Wannier-Stark functions $\psi_i(z)$ and
$\psi_j(z)$, and $f_{i,{\bf k}}$ and $f_{j,{\bf k}}$ are the nonequilibrium
distribution functions in subbands $i$ and $j$. The factor 2 comes from
summing over  the spin index.
Assuming parabolic bands and replacing the 
$\delta$ function with a Lorentzian to model the broadening of
the levels we obtain
\begin{equation}
{\rm Im}[\chi(\omega)] \; = \; \frac{\pi}{\epsilon_0 d}
\sum_{{ij \atop (E_j > E_i)}}|d_{ij}|^2 (n_i - n_j)\, \mathcal{L}_{ij}(\omega),
\label{eq.chi2lev}
\end{equation}
with the Lorentzian $\mathcal{L}_{ij}(\omega) = 
(\Gamma_{ij}/2\pi)/[(\hbar\omega - \Delta E_{ji})^2
+ (\Gamma_{ij}/2)^2]$, where $ \Gamma_{ij} = \Gamma_i + \Gamma_j$, and
$\Delta E_{ji} = E_j - E_i$ (these parameters are defined 
in Sec.~\ref{subsec.neqstate}). 
$n_i$ ($n_j$) is the 2D carrier density in subband $i$ ($j$).
In Eq.~(\ref{eq.chi2lev}), the contributions from all transitions with
$E_j > E_i$ are included. Using Eq.~(\ref{eq.chi2lev}) with 
Eq.~(\ref{eq.gaincof}) gives us an estimate of the gain coefficient 
$g_{WS}(\omega)$
within the Wannier-Stark picture. 

Disadvantages of this simpler approach,
compared to the approach described in Sec.~\ref{subsec.gaintheory},
are, firstly, that the Wannier-Stark levels become increasingly delocalized
at low bias and an increasingly large number of basis states must be used to
obtain a good approximation to the energy eigenstates as the applied
voltage decreases. Thus, this approach becomes impracticable to apply
at low bias. In addition, the function ${\rm Im}[G^{\rm ret}(E)]$ is
not always a simple Lorentzian as seen in Fig.~\ref{fig.gretek}(b)
for each Wannier-Stark level, but may, for some levels,
have additional peaks or shoulders indicating that these levels
do not approximate well a single eigenstate of the system but
are superpositions of the eigenstates. This behavior occurs mainly
at low bias, but can also occur when two Wannier-Stark levels are very close
in energy which gives a double-peaked structure to 
${\rm Im}[G^{\rm ret}(E)]$, e.g.,
for the two levels in panel (b) in Fig.~\ref{fig.wspage}.
In these cases, it is not possible to use this simple model to estimate
the gain.
Another drawback of this model is that by only making use of the
diagonal elements of the Green's functions, 
$G_{{\rm WS},ii}^<$ and $G_{{\rm WS},ii}^{\rm ret}$, we have
discarded information on quantum effects that are contained in the
offdiagonal elements. 

\begin{figure}

\includegraphics[height=10cm,keepaspectratio]{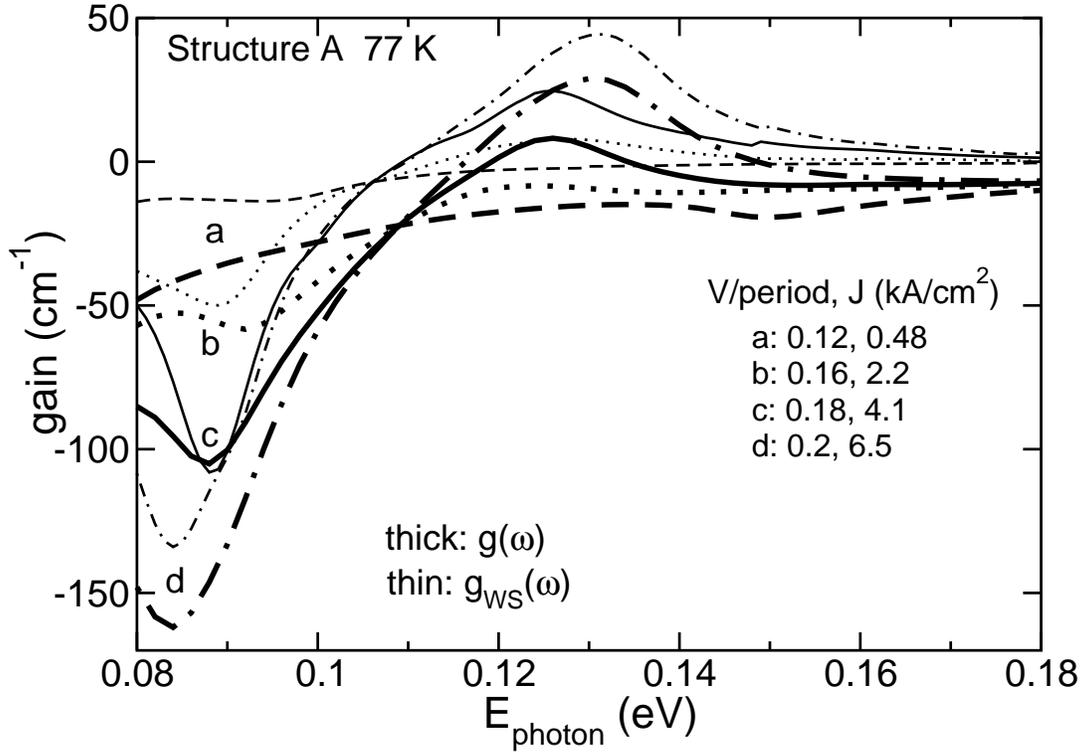}
\caption{
Comparison of $g(\omega)$ and $g_{\rm WS}(\omega)$, with different applied
bias, for structure A. a: dashed lines, b: dotted lines, c: solid lines,
d: dot-dashed lines.
\label{fig.gainNGF2comp}
}
\end{figure}

In Fig.~\ref{fig.gainNGF2comp}, we compare
the gain coefficient $g(\omega)$ from the  NGF theory with
the gain coefficient $g_{WS}(\omega)$ from the simple  model in the 
Wannier-Stark basis.
Spectra are shown for a bias of 0.12 -- 0.2 V/period.
We find that the simpler model reproduces well the main features
of the  NGF theory.
As in the previous figure, we see that for the simpler theory
in Fig.~\ref{fig.gainNGF2comp}, the gain curves are again 
offset to higher values compared to the more rigorous theory. 

\begin{figure}

\includegraphics[height=10cm,keepaspectratio]{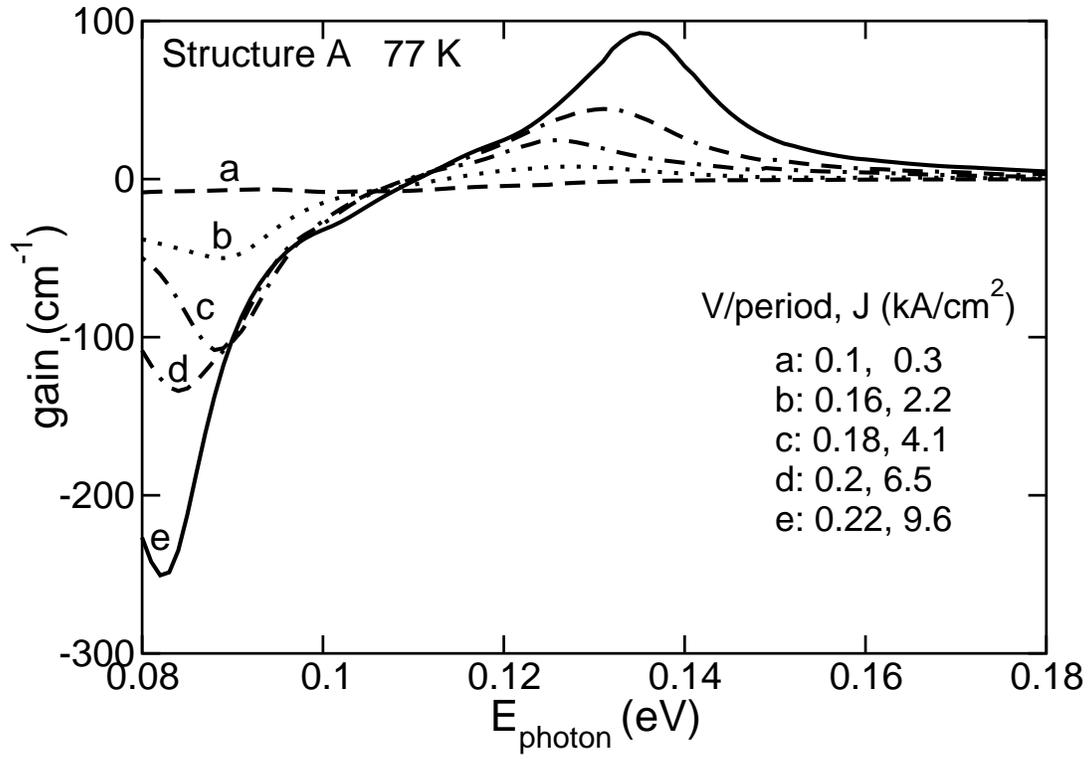}
\caption{Gain spectra $g_{\rm WS}(\omega)$ evaluated in
Wannier-Stark basis for structure A.
\label{fig.wsgain}
}
\end{figure}

We consider now the origin of the different features in the spectra.
This can  be most simply understood if we look at the gain spectra 
obtained  in the Wannier-Stark basis. For clarity, only these
spectra are shown again in Fig.~\ref{fig.wsgain}, and for a wider bias range,
0.1  to 0.22 V/period, than seen in Fig.~\ref{fig.gainNGF2comp}.
These spectra contain
transitions between all possible pairs of levels with energy
separation $\Delta E_{ji}$ in the energy range shown on the $x$ axis.
At the lowest bias shown, i.e., 0.1 V/period, we see that
the structure is almost transparent with only a very small
absorption for the energy range shown. As the applied voltage increases,
absorption increases in the range 0.08 to 0.11 eV. Simultaneously,
gain appears and increases in the range 0.11 to 0.16 eV. The gain
peak shifts to higher energies as the voltage increases.

\begin{figure}

\includegraphics[height=10cm,keepaspectratio]{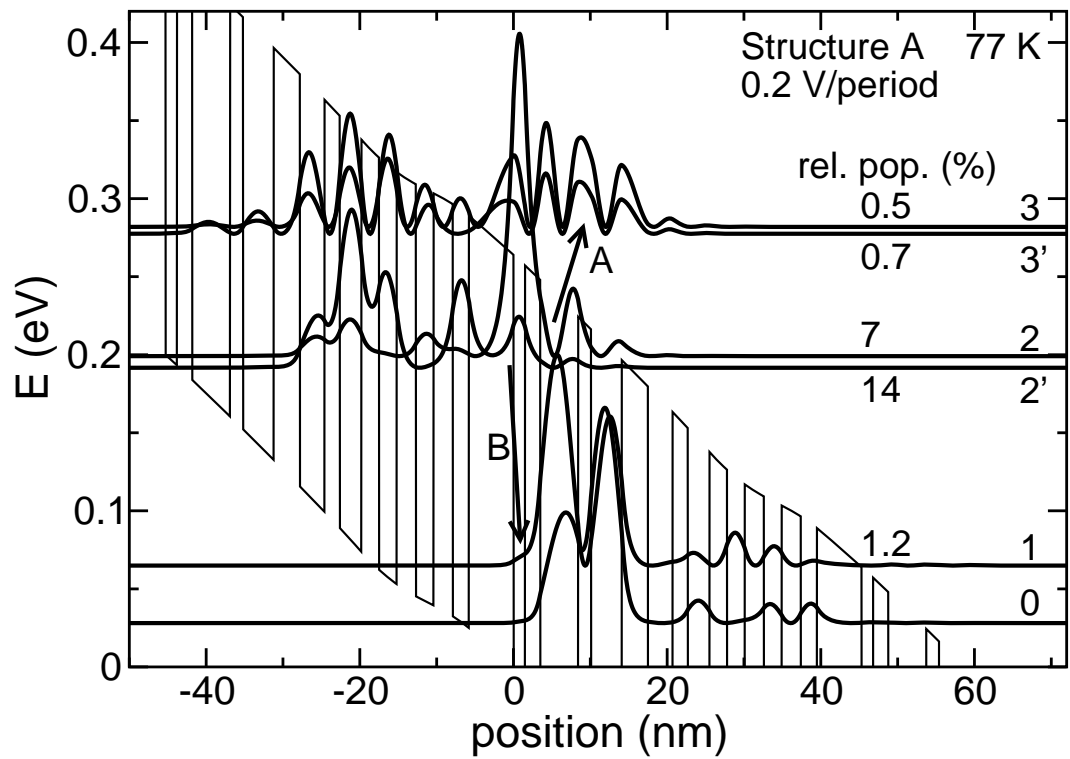}
\caption{Wannier-Stark levels in structure A at 0.2 V/period.
\label{fig.wslevel}
}
\end{figure}

To explain these features, 
we consider the curve at 0.2 V/period in Fig.~\ref{fig.wsgain}.
In Fig.~\ref{fig.wslevel}, we show the main Wannier-Stark energy
levels involved in the transitions which give rise to the absorption
and gain features seen at this bias.
From an inspection of the transitions 
contributing to the sum in Eq.~(\ref{eq.chi2lev}), 
we find as expected that the gain feature 
originates mainly from the transition (marked B in the figure)
between the upper laser level (labeled 2) and the
lower laser level (1). 
There is also some contribution
to the gain from transitions between an injector level (2')
and the lower laser level (1), and between other injector levels 
and the upper or lower laser level
(not shown). The absorption feature below 0.1 eV is due mainly 
to transitions (marked A in the figure)
from the upper laser level (2) to levels in the continuum
(3 and 3'), with additional contributions also from transitions
between injector levels. The relative population in each level
is also shown in the figure, and levels 2 and 2' have a greater
fraction of the population compared to levels 1, 3, and 3'. This
supports the attribution of gain to transition B, and absorption
to transition A. The largest fraction of the population $\sim$ 70\%
is in the lowest injector level (not shown in the figure). Level 0
contains around 2\% of the population.

When we consider the relative populations in these levels for different
applied voltages (in the range 0.1 -- 0.22 V/period as shown in 
Fig.~\ref{fig.wsgain}), we find that the population in the lower laser
level remains at around 1\% at all voltages. The population in the
upper laser level, however, changes from around 0.2\% at 0.1 V/period
to over 10\% at 0.22 V/period. Hence, there is a transition to a population
inversion between the two levels as the bias increases. The population in
level 0 ranges from around 20\% at 0.1 V/period to around 2 to 3\%
above 0.16 V/period. The largest proportion
of the population remains in the injector region, particularly in the lowest
injector level which forms a reservoir of around 60 to 70\% of the carriers
at all applied voltages. 

\begin{figure}

\includegraphics[height=10cm,keepaspectratio]{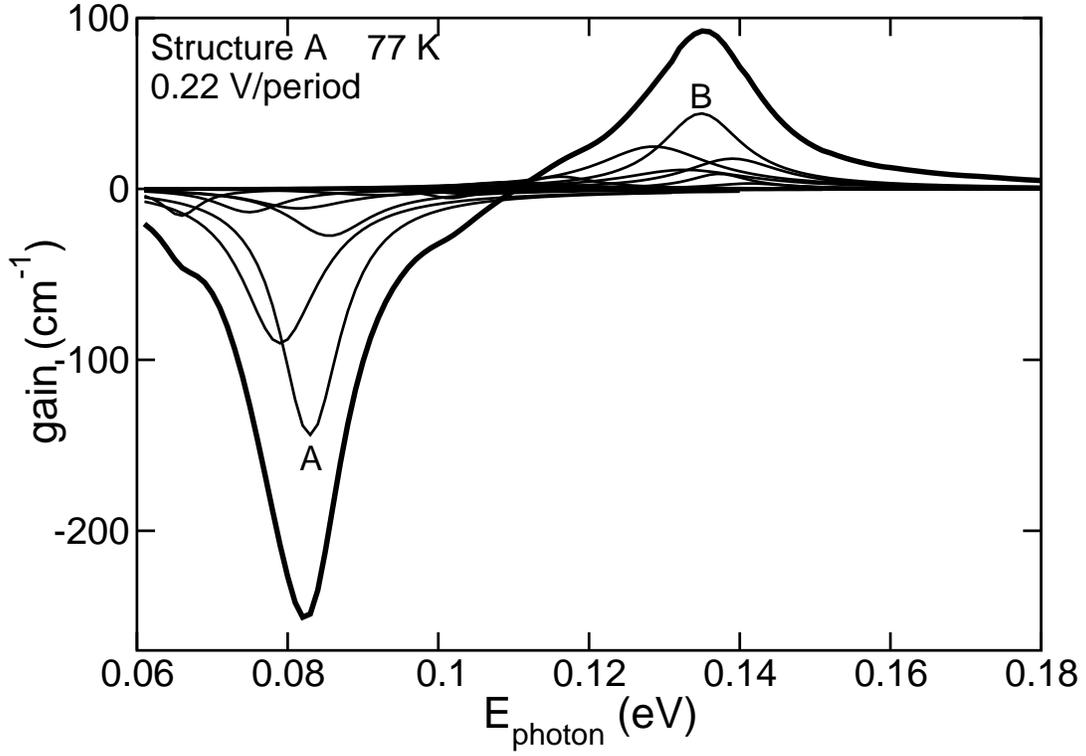}
\caption{Contributions of individual transitions (thin lines)
to gain curve $g_{\rm WS}(\omega)$ (thick line) at 0.22 V/period
for structure A
\label{fig.wsgainsep}
}
\end{figure}

We also note here that the broad gain or absorption features seen in
Fig.~\ref{fig.wsgain} are not due to transitions between a single
pair of levels but contain contributions from many transitions. 
For example, in Fig.~\ref{fig.wsgainsep},
we show the different Lorentzian contributions (thin lines) to the gain curve 
(thick line) at 0.22 V/period. Although the main contribution to the
absorption feature (thin line labeled A) originates from the upper laser
level -- continuum transition, and the main contribution to the
gain feature (thin line labeled B) arises from the upper -- lower laser
level transition, there is also a substantial contribution from
other (energetically) neighboring  transitions. The figure also shows that the
inhomogeneous broadening due to the contributions from different
transitions, and the broadening due to scattering, observed in the linewidth
of each individual Lorentzian, are comparable in size. At higher applied bias
the number of contributing transitions decreases, e.g, at 0.3 V/period, the
contributions to the gain feature come mainly from only two transitions.

\begin{figure}

\includegraphics[height=10cm,keepaspectratio]{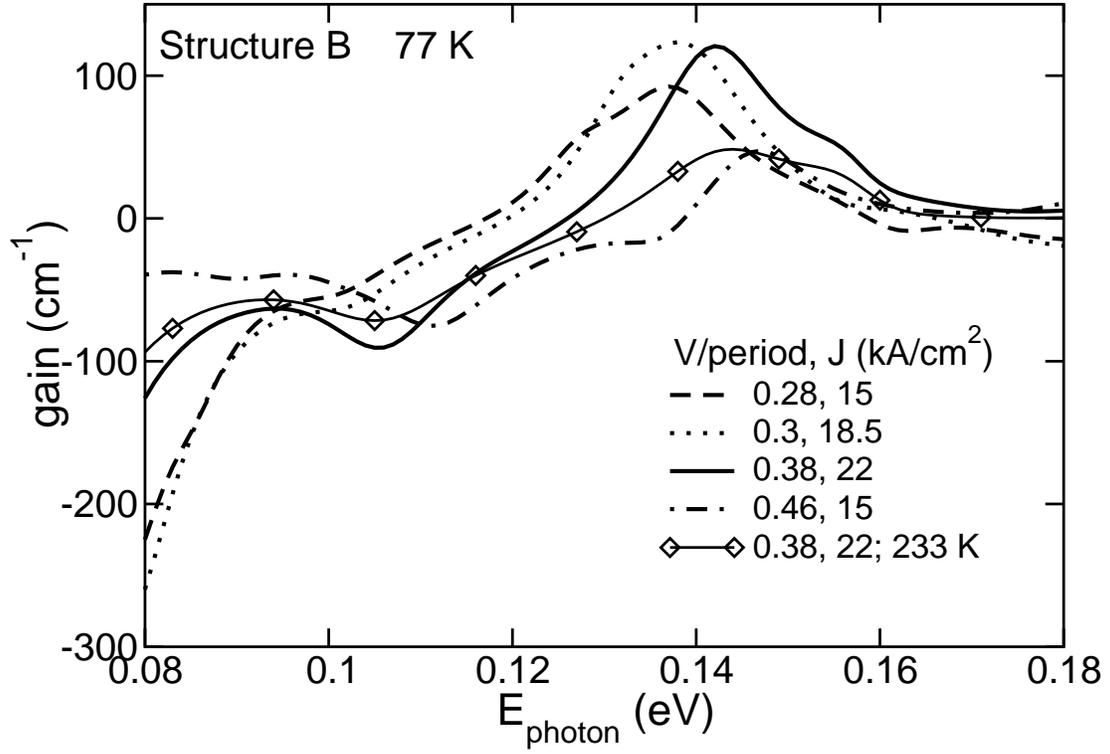}
\caption{Gain curves $g_{\rm WS}(\omega)$ at 77 K for structure B,
for different applied voltages. Thin solid line with symbols is at 233 K.
\label{fig.pgain}
}
\end{figure}

Finally, in the last set of results, we consider structure B. Experimental
data showing the light output vs. current density is shown in Fig.~4 of
Ref.~\onlinecite{Pag01}. At 77 K, the light output is seen to increase with
increasing current density until it reaches a maximum at around 22 kAcm$^{-2}$,
corresponding to the peak in the I-V characteristic seen in 
Fig.~\ref{fig.viexp}(b).
The light output goes to zero for higher current densities. 
A similar behavior is seen at 233 K 
but with a much reduced light output.
Figure~\ref{fig.pgain} shows the gain coefficient $g_{WS}(\omega)$ 
calculated at 77 K for structure B at different applied voltages 
around the peak of the I-V characteristic 
shown in Fig.~\ref{fig.viexp}(b). The gain feature increases as the
applied voltage or current density increases, reaching a maximum
around the peak of the I-V characteristic, and then decreases beyond
this point. A similar behavior is also seen at 233 K except that
the gain is reduced at the higher temperature. For comparison,
a gain curve calculated at 233 K, at the peak current density 22 kAcm$^{-2}$,
is also shown in Fig.~\ref{fig.pgain}. The gain feature in this curve
is much smaller than the gain feature in the corresponding curve (at
22 kAcm$^{-2}$) at 77 K. Hence, the behavior of the calculated gain curves 
correlate well with the light output vs. current density curves measured
in the experiment.

\section{Conclusion}

We have applied an NGF theory to obtain a description of the nonequilibrium
stationary state of  QC structures under an applied bias. Using this
information, we evaluate the current-voltage characteristic of
example QC structures reported in the literature. The theoretical
results are quantitatively close to experimental I-V data, and reproduce
well the trends seen in the data. In addition,
we determine two contributions to the current density.
The first contribution $J_o$ is driven by $\hat{H}_o$ which is the
Hamiltonian for the superlattice potential with applied bias.
The other contribution $J^\prime_{\rm scatt}$
is driven by $\hat{H}_{\rm scatt} + \hat{H}_{\rm MF}$ which
describes the scattering processes in the structure.
We find that, in contrast to simple superlattice structures, 
$J^\prime_{\rm scatt}$
is the main contribution to the current in the QC structures we consider.

In addition, we have extended the theory  to determine the linear response of
the nonequilibrium stationary state of these structures to a small
applied optical perturbation. This enables us to evaluate the linear
susceptibility and hence the gain or absorption spectra of the structure.
We compare the spectra obtained using a more rigorous NGF theory in
which the changes $\delta {\bf G}$, $\delta {\mathbf \Sigma}$,
and $\delta J_{\rm scatt}$ due to the optical perturbation are considered,
to simpler models in which (i) only  $\delta {\bf G}$ is considered,
or (ii) by summing over transitions in a simple two-band model
(summing over different pairs of bands) with Lorentzian broadened
levels. We find that the simpler models result in spectra which 
are offset to higher values than in the more rigorous theory.
We have also made a detailed analysis of the origin of the
different gain and absorption features in the spectra, and of
the redistribution of population within the 
Wannier-Stark levels as the applied voltage changes.
The gain and absorption features correlate well with the
distribution of population within these levels, which is determined
from the NGF theory.

\begin{acknowledgments}
The authors thank the Deutsche Forschungsgemeinschaft for financial support
through Grant No. FOR394. Discussions with A. Amann, H. Page, J. Schlesner, 
and M. Woerner are gratefully acknowledged.
\end{acknowledgments}

\appendix

\section{}
\label{app.basis}
The construction of the Wannier function basis used in the
calculations is described in this appendix. 
As the first step, we solve the one-dimensional Schr\"{o}dinger equation 
\begin{equation}
\bigg[-\frac{\hbar^2}{2m_e(z)}\frac{d^2}{dz^2} + V(z)\bigg]\psi(z) = 
E \psi(z),
\label{eq.schroe}
\end{equation}
for the envelope functions $\psi(z)$. 
The spatially dependent superlattice potential $V(z)$ and the effective mass
$m_e(z)$ are assumed constant in each semiconductor layer, 
i.e., $V(z) = V_o$ and $m_e(z) = m_b$ in the barriers,
and $V(z) = 0$ and $m_e(z) = m_w$ in the wells. Equation (\ref{eq.schroe}) can
be solved with a transfer matrix method (for a textbook discussion 
see, e.g., Ref.~\onlinecite{Yu99}). 
In this approach, the envelope function in
a semiconductor layer $j$ is written as
\begin{equation}
\psi_j(z) = A_j e^{i k_j(E) (z - z_j)} + B_j e^{-i k_j(E) (z - z_j)},
\label{eq.envfunc}
\end{equation}
where $z_j$ labels the position of interface $j$, and 
$k_j(E) = \sqrt{2 m_j (E - V_j)}/\hbar$ with $m_j$ and $V_j$ the mass and potential
in that layer. Applying continuity conditions
\begin{eqnarray}
\psi_j & = & \psi_{j+1}, \\
\frac{1}{m_j}\frac{d\psi_j}{dz} & = & \frac{1}{m_{j+1}}\frac{d\psi_{j+1}}{dz},
\end{eqnarray}
at the interface $j+1$, gives
\begin{equation}
{A_{j+1} \choose B_{j+1}} = \mathcal{M}_j(E) {A_j \choose B_j},
\label{eq.ab}
\end{equation}
with 
\begin{equation}
\mathcal{M}_j(E) = \frac{1}{2}\Bigg( 
\begin{array}{cc}
\big( 1 + \frac{m_{j+1}k_j}{m_j k_{j+1}} \big) e^{ik_j(z_{j+1} - z_j)} &
\big( 1 - \frac{m_{j+1}k_j}{m_j k_{j+1}} \big) e^{-ik_j(z_{j+1} - z_j)} \\
\big( 1 - \frac{m_{j+1}k_j}{m_j k_{j+1}} \big) e^{ik_j(z_{j+1} - z_j)} &
\big( 1 + \frac{m_{j+1}k_j}{m_j k_{j+1}} \big) e^{-ik_j(z_{j+1} - z_j)} \\
\end{array}
 \Bigg).
\end{equation}
If a single period $d$ of the structure contains $M$ semiconductor layers, the
Bloch condition $\psi_q(z+d) = e^{iqd}\psi_q(z)$ implies
\begin{equation}
{A_{M+1} \choose B_{M+1}} = \prod_{j=1}^M \mathcal{M}_j(E) {A_1 \choose B_1} 
= e^{iqd}{A_1 \choose B_1}.
\label{eq.blochcond}
\end{equation}
For a given value of $q$, only certain values
of $E$ allow the solution of Eq.~(\ref{eq.blochcond}), and this defines
the miniband structure $E^\nu(q)$. For each $q$ value, we determine 
$E^\nu(q)$ numerically by looking for the zeroes of the 
determinant of the matrix product $\prod_{j=1}^M \mathcal{M}_j(E)$.
These zeroes were found by stepping through the energy $E$, and comparing the
signs of the determinant for two consecutive values of $E$ (separated by $\Delta E$).
When these signs are opposite this sets the first coarse bracketing 
of the zero position. This position is then further refined by halving
this interval and successive intervals up to a hundred times, while comparing
the determinant signs of the interval endpoints at each iteration. 
Once $E^\nu(q)$ is determined,
we can also obtain $A^\nu(q) = A_1$ and $B^\nu(q) = B_1$, and the
Bloch functions $\psi^\nu_q(z)$,
from Eqs.~(\ref{eq.envfunc}), (\ref{eq.ab}), and (\ref{eq.blochcond}). 
In the calculations reported here, $E^\nu(q)$ was evaluated for 500 $q$ points
for each miniband, and $\Delta E$ was set to 5 meV. The Bloch functions were
evaluated on a position grid with 1500 points per period. In structure A (B),
there are eight (nine) minibands below the conduction-band offset.
For the calculations here, we included one miniband above the continuum,
so the band index $\nu$ runs from 1 to 9 (10) for structure A (B).

In the next step, we construct the Wannier functions 
(associated with miniband $\nu$)
\begin{equation}
\psi^W_\nu(z - nd) = \sqrt{\frac{d}{2\pi}}
\int^{\pi/d}_{-\pi/d} dq\; e^{-inqd}\, \psi_q^\nu(z)
\end{equation} 
from a superposition of the  Bloch functions in miniband $\nu$. 
The Wannier functions constructed by this superposition are not
unique, and can be very different depending
on the phase of the Bloch functions which can be chosen arbitrarily for
each value of $q$. Ideally, we would like the Wannier functions 
to be as spatially localized as possible 
to reduce the interaction matrix elements between
Wannier functions in different periods. As a first step to construct
Wannier functions with this property, we first fix
the phase of the Bloch functions such that these functions
are real at some arbitrary position $x_s$. Different
values of $x_s$ can be tested, and the resulting Wannier 
functions checked to see if they are fairly well-localized, 
e.g., within one period of the structure. We then express
the position operator $\hat{z}$ in the basis made up of this
initial set of Wannier functions. 
Finally, we diagonalize the
resulting matrix representation of $\hat{z}$, and the resulting
eigenfunctions give us the required set of Wannier functions associated
with a given miniband. This process is repeated for each miniband.

The Wannier states associated with
a given miniband are degenerate, and their energy expectation values
lie at the center of the miniband. Their wave functions are
spatially displaced from each other, with a separation given by
the period of the structure, i.e, there is one Wannier state
in each period of the structure.
To set up the matrix representation of $\hat{z}$ 
for a given miniband, we 
use Wannier states in 11 neighboring periods.
The large number of periods used in this construction 
is necessary to improve the numerical accuracy of the result.

In the transport and gain calculations, however, 
we consider only couplings and matrix 
elements between Wannier functions in the same period, and 
nearest-neighbor periods. Hence, the matrices representing
the Green's functions and self-energies
are constructed with 27 basis states (for structure A)
from three periods.

\section{}
\label{app.selfe}

This appendix describes in more detail the evaluation of the
self-energies and scattering matrix elements in
Eqs.~(\ref{eq.sigrough}) -- (\ref{eq.sigless}). 
The self-energies contain summations over ${\bf k'}$ of the form
$\displaystyle
\sum_{\bf k'} |V_{\alpha\beta}({\bf k},{\bf k'})|^2 \{ G. F.\}$ where
$V_{\alpha\beta}$ is a generic matrix element representing interface
roughness, impurity, or phonon scattering, and $\{ G. F.\}$ represents a
Green's function $G_{\beta\beta,{\bf k'}}(E')$, with $E' = E$ or 
$E' = E \pm E_{\rm phon}$,
and including, 
for the case of phonon scattering, a phonon distribution
factor. Taking the summation to the continuous limit leads to
\begin{equation}
\sum_{\bf k'}|V_{\alpha\beta}({\bf k},{\bf k'})|^2 \{ G. F.\}
 \; = \; \frac{{\mathcal A}}{(2\pi)^2} \int_0^\infty dk' k' \int_0^{2\pi} 
 d\theta \;
|V_{\alpha\beta}(k,k',\theta)|^2 \{ G. F.\}.
\label{eq.selfe}
\end{equation}
We carry out the angle integration assuming the
Green's function term does not depend on the angle, and defining the
angle-integrated quantity $\displaystyle
[ f(k') ]_\theta = \int_0^{2\pi} d\theta  f(k',\theta)$
we obtain
\begin{eqnarray}
 \sum_{\bf k'}|V_{\alpha\beta}({\bf k},{\bf k'})|^2 \{ G. F.\} & = &
  \frac{{\mathcal A}}{(2\pi)^2} \int dk' k'\;
\Big[ |V_{\alpha\beta}( k,k')|^2 \Big]_\theta \,\{ G. F.\} \nonumber\\
& = & 
\frac{{\mathcal A}}{(2\pi)^2} \frac{m_e}{\hbar^2}\int dE_{k'} \;
 \Big[ |V_{\alpha\beta}( k,k')|^2 \Big]_\theta \,\{ G. F.\}  \nonumber\\
& = & \frac{1}{2\pi}\frac{{\mathcal A}\rho_o}{2}\, 
 \Big[ |V_{\alpha\beta}( k_{\rm typ},k'_{\rm typ})|^2 \Big]_\theta 
 \int dE_{k'}\; \{ G. F.\}.
\label{eq.vangint}
\end{eqnarray}
The integral over momentum $k'$ has been transformed to an integral over
energy $E_{k'}$, and the density of states 
$\rho_o = m_e/(\pi\hbar^2)$. 
In the last line of Eq.~(\ref{eq.vangint})
we assume the matrix element $V_{\alpha\beta}( k,k')$ is slowly varying
compared to the Green's function term and can be taken out of the integral.
It is evaluated at fixed momenta $k_{\rm typ}$ and $k'_{\rm typ}$, and
the choice of these momenta is described below.

From a comparison of the factors outside the integral, in the last line of
Eq.~(\ref{eq.vangint}), with a scattering or transition probability
rate derived from Fermi's golden rule we can define\cite{foot.fgrrate}
 a fictitious scattering
rate (in energy units)
\begin{equation}
\gamma_{\alpha\beta} \; = \; \frac{\mathcal{A}\,\rho_o}{2}\,
 \Big[ |V_{\alpha\beta}( k_{\rm typ},k'_{\rm typ})|^2 \Big]_\theta.
\label{eq.fgrrate}
\end{equation}
With this definition we can rewrite Eqs. (\ref{eq.selfe}) 
and (\ref{eq.vangint}) as
\begin{equation}
\sum_{\bf k'}|V_{\alpha\beta}({\bf k},{\bf k'})|^2 \{ G. F.\} \;
= \; \frac{\gamma_{\alpha\beta}}{2\pi}  \int dE_{k'}\; \{ G. F.\}.
\label{eq.selfegam}
\end{equation}
Estimates of $\gamma$ for different scattering processes are
given in Table \ref{tab.rates}.

\begin{figure}

\includegraphics[height=10cm,keepaspectratio]{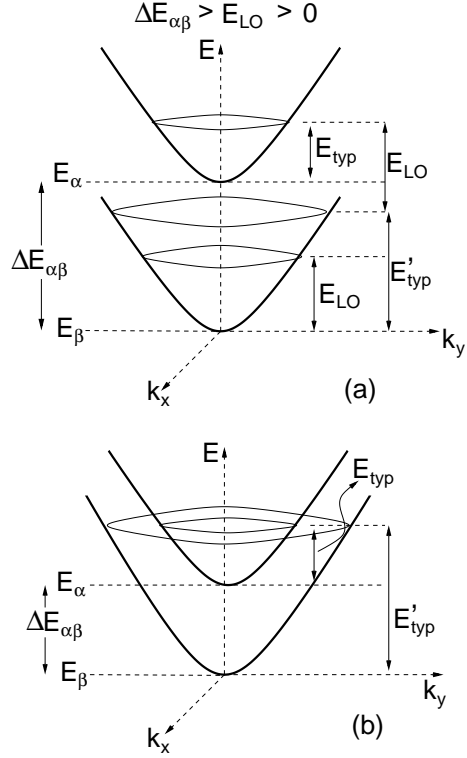}
\caption{Examples of $E_{\rm typ}$ and $E^{\prime}_{\rm typ}$
selection. (a) LO-phonon scattering for the case
$\Delta E_{\alpha\beta} \geq E_{\rm LO} > 0$ (requires
$E_{\rm typ} \geq 0$). Similar figures can be drawn for the cases
$E_{\rm LO} > \Delta E_{\alpha\beta} > 0$ 
(requires $E_{\rm typ} \geq E_{\rm LO} - \Delta E_{\alpha\beta}$), and
$\Delta E_{\alpha\beta} < 0$ 
(requires $E_{\rm typ} \geq |\Delta E_{\alpha\beta}|
+ E_{LO}$). For all three cases, $E^{\prime}_{\rm typ} = E_{\rm typ}
+ \Delta E_{\alpha\beta} - E_{LO}$. (b) interface roughness and impurity
scattering for $\Delta E_{\alpha\beta} > 0$.
\label{fig.bands}
}
\end{figure}

As stated after Eq.~(\ref{eq.sigless}), and earlier in this appendix,
we evaluate the scattering matrix elements
$V^{\rm phon/rough/imp}_{\alpha\beta}({\bf  k}, {\bf k'})$
using fixed momenta ${\bf k}_{\rm typ}$ and $ {\bf k'}_{\rm typ}$
(with the corresponding energies $E_{\rm typ}$ and  $E^{\prime}_{\rm typ}$) 
to accelerate the numerical computation. To fix $E_{\rm typ}$, we consider
the energy dependence of the scattering matrix elements. For
LO-phonon scattering, there is an energy threshold for phonon emission
because of energy conservation and the fixed phonon energy $E_{\rm LO} $.
The scattering matrix element is maximum at the emission threshold,
and decreases monotonically with increasing energy above this threshold.
To obtain an estimate of the scattering matrix element, which lies between
the higher values near threshold, and the lower values far above threshold,
we set $E_{\rm typ}$ one LO-phonon energy  
above the LO-phonon threshold.
We then fix $E^{\prime}_{\rm typ}
= E_{\rm typ} + \Delta E_{\alpha\beta} - E_{\rm LO}$, where 
$\Delta E_{\alpha\beta} =
[\hat{H}_o]_{\alpha\alpha} -[\hat{H}_o]_{\beta\beta}$ 
(see Fig.~\ref{fig.bands}).
To test the sensitivity of the results to
the value chosen for $E_{\rm typ}$, we have carried out runs 
with other values of
$E_{\rm typ}$, e.g., at threshold, $\frac{1}{2}E_{\rm LO}$ above threshold, 
and $2 E_{\rm LO}$ above threshold. The calculated current density 
changes by at most $\sim 9$\% (at some bias points with 
$E_{\rm typ}$ at threshold) 
but in most cases the change is around 5\% or much less (1 -- 2\%).
This tends to support the assumption that, for the LO-phonon process,
the results are not very sensitive to the specific value of $E_{\rm typ}$.
For impurity and interface roughness scattering, we set 
$E_{\rm typ} = 15$~meV, and $E^{\prime}_{\rm typ}
= E_{\rm typ} + \Delta E_{\alpha\beta}$.
Test runs were also carried out for 
$E_{\rm typ} = 1$, 7, and 30 meV. 
The results are more sensitive to the value of $E_{\rm typ}$.
For  $E_{\rm typ} = 7$ and 30 meV, the calculated current density changed 
by at most ~15\%.
For $E_{\rm typ} = 1$ meV, near the bottom of the subband, 
the difference was much larger, ranging from 10 to 50\%.
The value $E_{\rm typ} = 15$ meV was chosen as a value that lies near
the centre of the distributions in each subband, to give
an estimate of the average scattering 
matrix element.

With the momenta fixed, the scattering matrix elements can
be taken outside the integrals in Eqs.~(\ref{eq.sigrough}) -- 
(\ref{eq.sigless}),
and the self-energies depend only on integrals of the form 
[as shown in  Eq. (\ref{eq.vangint})]
\begin{equation}
G^{\rm ret}_{\alpha\alpha}(E) \; =\; \int_0^\infty dE_{k'}\,
 G_{\alpha\alpha,{\bf k'}}(E) ,
\label{eq.gretcomp}
\end{equation}
and
\begin{equation}
{\rm Im}[G^<_{\alpha\alpha}(E)] \; =\; \int_0^\infty dE_{k'}\,
{\rm Im}[G^<_{\alpha\alpha,{\bf k'}}(E)].
\end{equation}
Note that the diagonal elements of ${\bf G}^<$ are pure imaginary.
A problem arises in evaluating $G^{\rm ret}_{\alpha\alpha}(E)$ when
a momentum-independent scattering matrix element is used because
this leads to a divergence in the integral as $E_{k'}\rightarrow \infty$.
To deal with this problem, a cutoff energy for the upper limit of the
integral is used in the numerical integration.
The following subsections give more detail concerning the evaluation of
the matrix elements for the different scattering processes (see, also,
Refs. \onlinecite{Lak97}, \onlinecite{Che93}, \onlinecite{Rob93}).

\subsection{Interface roughness and impurity scattering}
\label{subsec.rough}

To derive matrix elements for interface roughness scattering we consider
an interface $j$ located at $z = z_j$ with thickness fluctuations
$\xi_j({\bf r})$  of the order of one monolayer. 
We assume correlation functions for the fluctuations given by
\begin{eqnarray}
\langle \xi_j({\bf r})\rangle &=&0
\label{eq.av1}\\
\langle \xi_j({\bf r})\xi_{j'}({\bf r'}) \rangle&=&
\delta_{j,j'}
\tilde{f}(|{\bf r}-{\bf r'}|)\quad \mbox{with}\quad
\tilde{f}(r)=
\eta^2\exp\left(-\frac{r}{\lambda}\right)\label{eq.av3}
\end{eqnarray}
$\eta$ denotes the root-mean-square of the roughness height and
$\lambda$ is a typical island size. Correlations between neighboring
interfaces are neglected. The angle brackets [as stated after
Eq.~(\ref{eq.sigrough})] denote an average over different distributions
of thickness fluctuations.

The Hamiltonian for interface roughness scattering is written as
\begin{equation}
\hat{H}_{\rm rough}=
\sum_{{{\bf k},{\bf p}\atop m\mu,n\nu}}\left[
V^{\rm rough}_{m\mu,n\nu}({\bf p})
\hat{a}_{m,\mu}^{\dag}({\bf k}+{\bf p})
\hat{a}_{n,\nu}({\bf k})+ {\rm H.c.}\right],
\end{equation}
with the matrix element
\begin{equation}
V^{\rm rough}_{m\mu,n\nu}({\bf p})=\sum_j\frac{1}{A}
\int d^2r \,e^{-i{\bf p}\cdot{\bf r}}\xi_j({\bf r})
\Delta E_c\psi^*_\mu(z_j-md)\psi_{\nu}(z_j-nd),
\end{equation}
where $\Delta E_c$ is the band offset.

Within the self-consistent Born approximation, the self-energy contribution
from interface roughness scattering is written as
\begin{equation}
\Sigma_{\alpha_1\alpha_2,{\bf k}}^{<,{\rm rough}}(E)
=\sum_{\beta\beta',{\bf k'}}
\langle V^{\rm rough}_{\alpha_1\beta}({\bf k}-{\bf k'})
V^{\rm rough}_{\beta'\alpha_2}({\bf k}'-{\bf k})
\rangle G^{<,{\rm rough}}_{\beta\beta',{\bf k'}}(E).
\label{eq.gensigrough}
\end{equation}
This equation is more general than Eq.~(\ref{eq.sigrough}) since it
includes the offdiagonal contributions. Now we assume that the diagonal
parts of $G^{<,{\rm rough}}_{\beta\beta',{\bf k'}}$ dominate, and we keep
only the terms $\beta = \beta'$ in the summation.
Then we obtain for the matrix element term 
\begin{eqnarray}
\langle V^{\rm rough}_{\alpha_1\beta}(-{\bf p}) 
V^{\rm rough}_{\beta\alpha_2}({\bf p})\rangle& = & 
\langle V^{\rm rough}_{m_1\mu_1,n\nu}(-{\bf p}) 
V^{\rm rough}_{n\nu,m_2\mu_2}({\bf p})\rangle
\nonumber\\
& = &
\frac{(\Delta E_c)^2}{\mathcal{A}^2}
\Big\langle \sum_{j'}
\int d^2r_1 \, e^{i{\bf p}\cdot{\bf r}_1 }\xi_{j'}({\bf r}_1)
\psi_{\mu_1}^*(z_{j'}-m_1d)\psi_{\nu}(z_{j'}-nd)\nonumber\\
& &\quad\times\sum_{j}  
\int d^2r_2\,e^{-i{\bf p}\cdot{\bf r}_2}\xi_j({\bf r}_2)
\psi_\nu^*(z_j-nd)\psi_{\mu_2}(z_j-m_2d)
\Big\rangle\nonumber\\
& = & \frac{(\Delta E_c)^2}{\mathcal{A}^2} \int d^2r_1\int d^2r_2 \,
e^{i{\bf p}\cdot({\bf r}_1 - {\bf r}_2)} 
\tilde{f}(|{\bf r}_1-{\bf r}_2|)\nonumber\\
& &\quad\times \sum_{j} \psi_{\mu_1}^*(z_{j}-m_1d)
|\psi_{\nu}(z_{j}-nd) |^2 \psi_{\mu_2}(z_j-m_2d) \nonumber \\
& = & \frac{(\Delta E_c)^2}{\mathcal{A}^2}
 \int d^2r e^{i{\bf p}\cdot{\bf r}} \tilde{f}(|{\bf r}|)\nonumber\\
& &\quad\times\sum_{j}\psi_{\mu_1}^*(z_{j}-m_1d)
|\psi_{\nu}(z_{j}-nd) |^2 \psi_{\mu_2}(z_j-m_2d),
\end{eqnarray}
where we have expanded the general indices
$\alpha_1,\alpha_2,\beta$ in terms of the period and Wannier level indices
[see after Eq.~(\ref{eq.gret})], and defined ${\bf p} = {\bf k'} - {\bf k}$.
Because of the orthogonality of the wave functions, and because the 
wave functions extend over many interfaces, the sum over $j$ tends to vanish 
for $m_1 \neq m_2$ and $\mu_1 \neq \mu_2$. 
This suggests replacing this term in the last line with
$F^{\mu_1\nu}_{m_1 -n}\delta_{m_1,m_2}\delta_{\mu_1,\mu_2}$ where
\begin{equation}
F^{\mu_1\nu}_{h}\;=\;
\sum_{j} |\psi_{\mu_1}^*(z_{j}-hd)|^2 |\psi_{\nu}(z_{j}) |^2.
\end{equation}
Applying Eq.~(\ref{eq.av3}) we obtain 
\begin{equation}
\langle V^{\rm rough}_{m\mu,n\nu}(-{\bf p}) 
V^{\rm rough}_{n\nu,m\mu}({\bf p})\rangle \; =\;
\frac{(\Delta E_c)^2\eta^2}{\rho_o\mathcal{A}E_{\lambda}}
\frac{1}{(1+E_p/E_{\lambda})^{3/2}}
F_{m-n}^{\mu\nu} \; =\; \mathcal{F}(E_p),
\label{eq.vmatrough}
\end{equation}
with $E_{\lambda}=\hbar^2/2m\lambda^2$ and
$E_p=\hbar^2p^2/2m = E_k + E_{k'} - 2\sqrt{E_k E_{k'}}\cos\theta$,
where $\theta$ is the angle between ${\bf k}$ and ${\bf k'}$. 
The subscript 1 from the indices $m_1$ and $\mu_1$ is dropped
for simplicity.
Substituting this latter result 
in Eq.~(\ref{eq.gensigrough}) gives Eq.~(\ref{eq.sigrough})
which contains only the diagonal terms in the self-energy and
the Green's function. We now follow the procedure outlined
in Eqs.~(\ref{eq.selfe}) -- (\ref{eq.vangint}) and take the
summation over ${\bf k'}$ in Eq.~(\ref{eq.sigrough}) to the
continuous limit.
We observe that both the self-energy and
the Green's function depend only on $E_k$ and $E_{k'}$, but
not on the angle $\theta$. Hence,
as shown in Eq.~(\ref{eq.vangint}),  the angle
integration over the matrix element can be carried out analytically,
 and we define the angle-integrated
quantity (Ref.~\onlinecite{Gra80}, 2.575):
\begin{eqnarray}
\Big[ \langle |V^{\rm rough}_{m\mu,n\nu}({\bf k} - {\bf k'})|^2 
\rangle \Big]_\theta
& = &\int_0^{2\pi} d\theta \;
\mathcal{F}(E_k + E_{k'} - 2\sqrt{E_k E_{k'}}\cos\theta)\nonumber\\
& = & \frac{(\Delta E_c)^2\eta^2}{\rho_o\mathcal{A}E_{\lambda}} 
F_{m-n}^{\mu\nu} \frac{4}{(a-b)\sqrt{a+b}}\, 
E\Bigg(\sqrt{\frac{2b}{a+b}}\Bigg),
\end{eqnarray}
with
\begin{equation}
a=1+\frac{E_k+E_{k'}}{E_{\lambda}}
\qquad\qquad b=2\frac{\sqrt{E_kE_{k'}}}{E_{\lambda}}.
\end{equation}
$E(x)$ is the complete elliptic integral of the second kind which
is of order  $\pi/2=E(0)>E(x)>E(1)=1$ (Ref. \onlinecite{Abr70}, 17.3). 
Therefore we set $E(x) \approx \pi/2$, and fixing $E_k = E_{\rm typ}$
and $E_{k'} = E_{\rm typ} + \Delta E_{\alpha\beta}$ as described earlier, 
we define the fictitious interface roughness
scattering rate [using Eq.~(\ref{eq.fgrrate})]
\begin{equation}
\gamma^{\rm rough}_{(m-n),\mu\nu} = 
\frac{\pi(\Delta E_c)^2\eta^2\sqrt{E_\lambda}F_{m-n}^{\mu\nu}}
{\left[E_{\lambda}+\left(\sqrt{E_{\rm typ}} -
\sqrt{E_{\rm typ}+\Delta E_{\alpha\beta}}\right)^2\right]
\sqrt{E_{\lambda}+\left(\sqrt{E_{\rm typ}}+
\sqrt{E_{\rm typ}+\Delta E_{\alpha\beta}}\right)^2}}.
\label{eq.gamrough}
\end{equation}

Impurity scattering is mediated through the (3D) Coulomb interaction 
$V \sim 1/q^2 \sim 1/|{\bf k} - {\bf k'}|^2$. 
Neglecting the angle-dependent term gives
$|V|^2 \sim 1/(E_k + E_k')^2$.
Fixing $E_k = E_{\rm typ}$
and $E_{k'} = E_{\rm typ} + \Delta E_{\alpha\beta}$ as before, and neglecting
factors of 2, we obtain an order-of-magnitude estimate of the 
impurity scattering rate
\begin{equation}
\gamma_{\alpha\beta}^{\rm imp} \; = \; 
\gamma_{\rm par}^{\rm imp}
\frac{\Big[ \int dz\,|\psi_\alpha^*(z)\psi_\beta(z)| \Big]^2}
{(1 + \Delta E_{\alpha\beta}/E_{\rm typ})^2}.
\label{eq.gamimp}
\end{equation}
The parameter $\gamma_{\rm par}^{\rm imp}$ (see Table \ref{tab.rates})
is estimated from calculations of impurity scattering in simple superlattice
structures.\cite{Wac97d}

\subsection{LO-phonon}
\label{subsec.lophonon}
To derive the electron--LO-phonon scattering matrix element we
start from the interaction Hamiltonian\cite{Fer89}
\begin{equation}
\hat{H}_{\rm lo} = \sum_{\bf Q}\,i \alpha({\bf Q}) 
[  e^{-i{\bf Q}\cdot {\bf R}}
\hat{b}^\dagger_{\bf Q} 
 -   e^{i{\bf Q}\cdot {\bf R}}
\hat{b}^{}_{\bf Q} ],
\end{equation}
with
\begin{equation}
\alpha({\bf Q}) = \biggl[\frac{e^2}{2 
 {\cal V}}\frac{\hbar\omega_{\rm lo}}{Q^2}
\biggl(\frac{1}{\epsilon_\infty} - \frac{1}{\epsilon_s}\biggr)
\biggr]^{1\over2}.
\end{equation}
${\bf Q} = ({\bf q},q_z)$
is the 3D phonon wave vector with the 2D in-plane component ${\bf q}$ and
the component  $q_z$ in the growth direction. Similarly, the 3D position
vector ${\bf R} = ({\bf r},z)$.
$\hat{b}^\dagger_{\bf Q}$
$(\hat{b}_{\bf Q})$  
is the phonon creation (annihilation) operator, $\hbar\omega_{\rm lo} =
E_{\rm lo}$ is the
LO-phonon energy, and $\mathcal{V}$ is the crystal volume.  $\epsilon_\infty$ 
and $\epsilon_s$ are the 
high-frequency and static absolute permittivities.

The matrix element
\begin{equation}
V^{\rm lo}_{\alpha\beta} \;=\;
 \langle \alpha | \hat{H}_{\rm lo} | \beta \rangle 
\;=\; \sum_{\bf Q}\,
i \alpha({\bf Q})\langle \alpha |e^{-i{\bf Q}\cdot {\bf r}}
\hat{b}^\dagger_{\bf Q} 
 -   e^{i{\bf Q}\cdot {\bf r}}
\hat{b}^{}_{\bf Q}| \beta \rangle,
\label{eq.fHi}
\end{equation}
is written with initial and final states, $| \beta \rangle$ and $| \alpha 
\rangle$, 
given by
\begin{eqnarray*}
| \beta \rangle & = &|\Psi_{{\bf k},\beta}( 
{\bf r},  z)\rangle
|n(\omega_{\rm lo})\rangle, \\
| \alpha \rangle & = &|\Psi_{{\bf k'},\alpha}(
 {\bf r}, z)\rangle
|n(\omega_{\rm lo}) \pm 1\rangle.
\end{eqnarray*}
The ket containing
 $n(\omega_{\rm lo})$ is a phonon number state. The upper (lower)
 sign in the ket
 $|n(\omega_{\rm lo}) \pm 1\rangle$ corresponds to phonon emission (absorption).
 The ket containing $\Psi$ gives the electron state, and
 the electron wave function
$ \Psi_{{\bf k},\beta}(
 {\bf r}, z) = {\cal A}^{-{1\over2}}
e^{i{\bf k}\cdot{\bf r}}
\psi_\beta( z).$
Evaluating Eq. (\ref{eq.fHi}) leads to
\begin{equation}
|V^{\rm lo}_{\alpha\beta}({\bf k},{\bf k'})|^2 \;=\;
|\langle \alpha | \hat{H}_{\rm lo} | \beta \rangle|^2 \; = \;
\sum_{\bf Q}\,\alpha^2({\bf Q}) [ n(\omega_{\rm lo}) + \hbox{$  1\over2$} 
\pm \hbox{$ 1\over2$}]|M_{\alpha\beta}(q_z)|^2
\delta_{{\bf k}',{\bf k}\mp {\bf q}},
\label{eq.mtx}
\end{equation}
with 
$$M_{\alpha\beta}(q_z) = \int_0^{L_w} dz\; e^{\mp i q_z z} 
\psi_\alpha^*(z)\psi_\beta(z).$$
$L_w$ is the distance in the $z$-direction over which the
wave functions $\psi(k_z,z)$ extend.
Converting the sum over {\bf Q} in Eq. (\ref{eq.mtx}) to
an integral, and evaluating the in-plane component 
${\bf q}$ with the help of the Kronecker
delta gives
\begin{equation}
 |V^{\rm lo}_{\alpha\beta}({\bf k},{\bf k'})|^2\; = \; [n(\omega_{\rm lo}) + 
\hbox{$  1\over2$} 
\pm \hbox{$ 1\over2$}]
\frac{E_{\rm lo} e^2}{4\pi\epsilon_p\mathcal{A}}
\int_{-\infty}^{\infty} dq_z \frac{|M_{\alpha\beta}(q_z)|^2}{q_z^2 + p^2},
\end{equation}
where $p = |{\bf k} - {\bf k'}|^2 = k^2 + {k'}^2 - 2k k' \cos\theta$. 

As described earlier in this appendix, the matrix element is 
evaluated within a self-energy integral, for
instance of the form:
$\Sigma({\bf k},E) \; = \; 
\sum_{\bf k'} |V_{\alpha\beta}({\bf k},{\bf k'})|^2
\{G.F\}$. Following again the procedure shown in 
Eqs.~(\ref{eq.selfe}) -- (\ref{eq.vangint}), we take the
summation over ${\bf k'}$ in the self-energy to the
continuous limit, and we define the angle-integrated quantity

\begin{eqnarray}
\Big[ \langle |V^{\rm lo}_{\alpha\beta}({\bf k},{\bf k'})|^2 \Big]_\theta 
& = & C \int_0^{2\pi} d\theta
\int_{-\infty}^{\infty} dq_z
\frac{|M_{\alpha\beta}(q_z)|^2}{q_z^2 + k^2 + {k'}^2 - 2k k'\cos\theta}
\nonumber\\
& = &  2\pi C
\int_{-\infty}^{\infty} dq_z
\frac{|M_{\alpha\beta}(q_z)|^2}{
\sqrt{(q_z^2 + k^2 + {k'}^2)^2 - 4k^2 {k'}^2}}
\nonumber\\
& = &\frac{C}{\rho_o} \int_{-\infty}^{\infty} dq_z
\frac{|M_{\alpha\beta}(q_z)|^2}{
\sqrt{(\frac{\hbar^2 q_z^2}{2 m} + E_k + E_{k'})^2 - 4 E_k E_{k'}}},
\end{eqnarray}
where $C =  E_{\rm lo} e^2[n(\omega_{\rm lo}) + \hbox{$  1\over2$} 
\pm \hbox{$ 1\over2$}] /(4\pi\epsilon_p\mathcal{A})$.

Using this result in Eq.~(\ref{eq.fgrrate}), and
fixing the energies $E_k = E_{\rm typ}$ and $E_{k'} = E'_{\rm typ}$ 
as described earlier, we can define the scattering rate
\begin{equation}
\gamma_{\alpha\beta}^{\rm lo} = [n(\omega_{\rm lo}) + \hbox{$  1\over2$} 
\pm \hbox{$ 1\over2$}]\frac{E_{\rm lo} e^2}{8\pi\epsilon_p}
\int_{-\infty}^{\infty} dq_z\frac{|M_{\alpha\beta}(q_z)|^2}{
\sqrt{(\frac{\hbar^2 q_z^2}{2 m} + E_{\rm typ} + E'_{\rm typ})^2 
- 4 E_{\rm typ} E'_{\rm typ}}}.
\label{eq.gamlo}
\end{equation}

\subsection{Acoustic phonon}
\label{subsec.acphonon}

The phonons are implemented as an artificial optical phonon with a phonon
energy $E_{\rm ac}$ which should be smaller than $k_B T$ and which should
not be commensurable with the optical phonon energy. The matrix element,
or equivalently, the fictitious scattering rate $\gamma_{\alpha\beta}^{\rm ac}$,
is set to
\begin{equation}
\gamma_{\alpha\beta}^{\rm ac} \; = \; \gamma_{\rm par}^{\rm ac} 
\Big[ \int dz\,|\psi_\alpha^*(z)\psi_\beta(z)| \Big]^2.
\label{eq.gamac}
\end{equation}

\section{}
\label{app.jscatt}
A derivation of the current contribution $J_{\rm scatt}$ 
[Eq.~(\ref{eq.jscatt})] is given here. We assume that  
the scattering Hamiltonian $\hat{H}_{\rm scatt}$ has the form
\begin{equation}
\displaystyle \sum_{{\alpha\beta \atop {\bf k, k'},s}}
\hat{O}_{\alpha{\bf k},\beta{\bf k'}}(t)
\hat{a}^\dagger_{\alpha{\bf k}s}(t)\hat{a}_{\beta{\bf k'}s}(t),
\end{equation}
where $\hat{O}_{\alpha\beta}(t)$ may be just a scalar time-independent
matrix element, i.e. 
$\hat{O}_{\alpha{\bf k},\beta{\bf k'}}(t) = V_{\alpha{\bf k},\beta{\bf k'}}$,
as in interface roughness or impurity scattering. Alternatively, as
in phonon scattering, it may be
a time-dependent operator containing the phonon operators 
$\hat{b}(t)$ and  $\hat{b}^{\dag}(t)$. Then we find
\begin{eqnarray}
J_{\rm scatt}(t) & = & \frac{e}{\mathcal{V}}\frac{i}{\hbar} 
                  \langle [\hat{H}_{\rm scatt},\hat{z}]\rangle\nonumber \\
& = & \frac{e}{\mathcal{V}}\frac{1}{\hbar} 
\sum_{{\alpha\beta\gamma \atop {\bf k, k'},s}} i
        \langle \hat{a}^\dagger_{\alpha{\bf k}s}(t) 
          [\hat{O}_{\alpha{\bf k},\gamma{\bf k'}}(t) z_{\gamma\beta} 
           \,-\, z_{\alpha\gamma}\hat{O}_{\gamma{\bf k},\beta{\bf k'}}(t) ] 
           \hat{a}_{\beta{\bf k'}s}(t) \rangle\, . \label{eq.jscattgen}
\end{eqnarray}
In the following, this expression will be expressed in terms
of Green's functions and self-energies. For this purpose, we
consider the contour-ordered
Green's function:
\begin{equation}
F^c_{{\alpha\beta\gamma}\atop {\bf k}, {\bf k'},s}(\tau_1,\tau_2) 
\;=\; -i\hat{T}_c\{ 
\langle \hat{a}_{\beta{\bf k'}s}(\tau_1) 
\hat{a}^\dagger_{\alpha {\bf k}s}(\tau_2) 
\hat{O}_{\alpha{\bf k},\gamma{\bf k'}}(\tau_2)
z_{\gamma\beta} \,-\, 
z_{\alpha\gamma}\hat{O}_{\gamma{\bf k},\beta{\bf k'}}(\tau_1)
\hat{a}_{\beta{\bf k'}s}(\tau_1) 
\hat{a}^\dagger_{\alpha {\bf k}s}(\tau_2)\rangle\},
\end{equation}
where $\hat{T}_c$ orders the times arguments $\tau$ on the complex contour
from $\tau=-\infty+i0^+$ to $\tau=-\infty-i0^+$ (see Ref.~\onlinecite{Hau96}).
By taking the lesser component  
$F^<_{\alpha\beta\gamma \atop {\bf k, k'},s}(t,t)$,
one obtains the summand in Eq.~(\ref{eq.jscattgen}).

Using the interaction picture with superscript $D$,
$F^c$ is now evaluated according to the
standard perturbation expansion of
\begin{eqnarray}
F^c_{\alpha\beta\gamma \atop {\bf k, k'},s}(\tau_1,\tau_2) & = & 
-i\hat{T}_c
\Bigg\{ 
\Bigg\langle
\exp\left(-\frac{i}{\hbar}  
\int d\tau \hat{H}^D(\tau)\right) 
\Big[
\hat{a}^D_{\beta{\bf k'}s}(\tau_1) \hat{a}^{D\dagger}_{\alpha{\bf k}s}(\tau_2) 
\hat{O}_{\alpha{\bf k},\gamma{\bf k'}}(\tau_2)
z_{\gamma\beta}
\nonumber\\
 & & \qquad\qquad\qquad
 \,-\, z_{\alpha\gamma}\hat{O}_{\gamma{\bf k},\beta{\bf k'}}(\tau_1)
\hat{a}^D_{\beta{\bf k'}s}(\tau_1) 
\hat{a}^{D\dagger}_{\alpha{\bf k}s}(\tau_2)\Big]
\Bigg\rangle 
\Bigg\}.
\label{eq.fc}
\end{eqnarray}
The exponential term can be expanded as
\begin{equation}
\exp\left(-\frac{i}{\hbar}  \int d\tau \hat{H}^D(\tau)\right)\approx
1-\frac{i}{\hbar}  \int d\tau
\sum_{\delta\epsilon\atop {\bf p, q},s'}
\hat{O}_{\delta{\bf p},\epsilon{\bf q}}(\tau) 
\hat{a}^{D\dagger}_{\delta{\bf p}s'}(\tau) 
\hat{a}^D_{\epsilon{\bf q}s'}(\tau). 
\end{equation}
Substituting this expansion into Eq.~(\ref{eq.fc}),
and noting that terms containing only a single $\hat{O}$
are zero after averaging, 
the lowest-order nonvanishing terms give:
\begin{eqnarray}
& &F^c_{\alpha\beta\gamma \atop {\bf k, k'},s}(\tau_1,\tau_2) \;\approx\; 
\frac{1}{\hbar} \sum_{\delta\epsilon} \int d\tau\,
\Big[ G^{c0}_{\beta{\bf k'}s,\delta{\bf k'}s}(\tau_1,\tau)
{O}_{\delta{\bf k'},\epsilon{\bf k}}(\tau)
G^{c0}_{\epsilon{\bf k}s,\alpha{\bf k}s}(\tau,\tau_2)
{O}_{\alpha{\bf k},\gamma{\bf k'}}(\tau_2)
z_{\gamma\beta} \nonumber\\ 
& & \qquad\qquad\qquad \qquad\qquad -z_{\alpha\gamma}
{O}_{\gamma{\bf k},\beta{\bf k'}}(\tau_1)
G^{c0}_{\beta{\bf k'}s,\delta{\bf k'}s}(\tau_1,\tau)
{O}_{\delta{\bf k'},\epsilon{\bf k}}(\tau)
G^{c0}_{\epsilon{\bf k}s,\alpha{\bf k}s}(\tau,\tau_2)\Big]
\label{eq.fcg}
\end{eqnarray}
with the bare Green's functions
$G^{c0}_{\alpha{\bf k}s,\beta{\bf k}s}(\tau_1,\tau)
=-i\hat{T}_c\{ \langle \hat{a}^D_{\alpha{\bf k}s}(\tau_1) 
\hat{a}^{D\dagger}_{\beta{\bf k}s}(\tau)\rangle\}$. Only Green's functions
diagonal in the momentum and spin indices are kept in Eq. (\ref{eq.fcg}).
In order to be consistent with the perturbation expansion in
the Green's functions, further terms are taken into account, which
replace the bare Green's functions by the full Green's functions. Then we 
find
\begin{eqnarray}
\sum_{\bf k, k'}
F^c_{\alpha\beta\gamma \atop {\bf k, k'},s}(\tau_1,\tau_2) & =  &
\frac{1}{\hbar} \sum_{\delta,{\bf k}} \int d\tau
\big[ G^c_{\beta{\bf k}s,\delta{\bf k}s}(\tau_1,\tau)
\Sigma^{c(\alpha,r)}_{\delta\gamma{\bf k}}(\tau,\tau_2)z_{\gamma\beta} 
\nonumber\\
& &\qquad\qquad\qquad\qquad\, - \, 
z_{\alpha\gamma}\Sigma^{c(\beta,l)}_{\gamma\delta{\bf k}}(\tau_1,\tau)
G^c_{\delta{\bf k}s,\alpha{\bf k}s}(\tau,\tau_2) \big]
\label{eq.fcdum}
\end{eqnarray}
where
\begin{equation}
\Sigma^{c(\alpha,r)}_{\delta\gamma,{\bf k}}(\tau,\tau_2)=
\sum_{\epsilon{\bf k'}}O_{\delta{\bf k},\epsilon{\bf k'}}(\tau)
G^{c}_{\epsilon{\bf k'}s,\alpha{\bf k'}s}(\tau,\tau_2)
O_{\alpha{\bf k'},\gamma{\bf k}}(\tau_2)
\end{equation}
denotes the part of the self-energy
which exhibits ${O}_{\alpha\gamma}$ on the right-hand side, and
\begin{equation}
\Sigma^{c(\beta,l)}_{\gamma\delta,{\bf k}}(\tau_1,\tau)=
\sum_{\epsilon{\bf k'}}O_{\gamma{\bf k},\beta{\bf k'}}(\tau_1)
G^{c}_{\beta{\bf k'}s,\epsilon{\bf k'}s}(\tau_1,\tau)
O_{\epsilon{\bf k'},\delta{\bf k}}(\tau)
\end{equation}
with ${O}_{\gamma\beta}$  on the left-hand side. In the second term in
Eq.~(\ref{eq.fcdum}) we have exchanged the dummy indices 
$\delta$ and $\epsilon$,
and in the first term we have exchanged ${\bf k}$ and ${\bf k'}$.

For diagonal self-energies, 
which depend on diagonal Green's functions (see Sec.~\ref{subsec.qteself}), 
one has $\Sigma^{c(\alpha,r/l)}_{\delta\gamma,{\bf k}} =
\delta_{\delta\gamma}\Sigma^{(\alpha)}_{\gamma\gamma,{\bf k}}$ where
$\Sigma^{(\alpha)}_{\gamma\gamma,{\bf k}}$ is defined after 
Eq.~(\ref{eq.jscatt}).
Using Langreth rules\cite{Lan76a} and changing to the energy 
representation leads to
Eq.~(\ref{eq.jscatt}) which is given again here for reference:
\begin{eqnarray}
J_{\rm scatt} & = & \frac{2e}{\hbar\mathcal{V}}\sum_{\alpha\beta\gamma}
                 \sum_{\bf k}
	      \int \frac{dE}{2\pi} 
	      \underbrace{\Big( G^<_{\beta\gamma,{\bf k}}(E)
	      \Sigma^{{\rm adv}(\alpha)}_{\gamma\gamma,{\bf k}}(E)\; +\;
	      G^{\rm ret}_{\beta\gamma,{\bf k}}(E) 
	      \Sigma^{<(\alpha)}_{\gamma\gamma,{\bf k}}(E)
	      \Big) z_{\gamma\beta}}_{\rm I} \nonumber\\
	      &   & \quad -\; \underbrace{z_{\alpha\gamma} \Big(
	      \Sigma^{<(\beta)}_{\gamma\gamma,{\bf k}}(E) 
	      G^{\rm adv}_{\gamma\alpha,{\bf k}}(E)
	      \; + \; 
	      \Sigma^{{\rm ret}(\beta)}_{\gamma\gamma,{\bf k}}(E)
	      G^<_{\gamma\alpha,{\bf k}}(E)
	      \Big)}_{\rm II}.
\label{eq.appjscatt}
\end{eqnarray}

We can obtain some insight into this expression for $J_{\rm scatt}$
if we consider only the diagonal terms of the above equation, 
i.e., we set $\gamma = \beta$ in part I, and $\gamma = \alpha$ in part II, 
to obtain
\begin{eqnarray}
\int dE \;{\rm I} & \longrightarrow & 
\int dE\; 
\Big(  \underbrace{
G^<_{\beta\beta}\,\Sigma^{{\rm adv}(\alpha)}_{\beta\beta}}_{\rm Ia}
\;+\; \underbrace{G^{\rm ret}_{\beta\beta}\,
\Sigma^{<(\alpha)}_{\beta\beta}}_{\rm Ib}
\Big)
\;z_{\beta\beta},
\label{eq.Idiag}
\\
\int dE\; {\rm II} & \longrightarrow & 
\int dE\; z_{\alpha\alpha}\; 
\Big(\underbrace{
\Sigma^{<(\beta)}_{\alpha\alpha}G^{\rm adv}_{\alpha\alpha}}_{\rm IIa}
\; + \;  \underbrace{
 \Sigma^{{\rm ret}(\beta)}_{\alpha\alpha}G^<_{\alpha\alpha}}_{\rm IIb}\Big),
\label{eq.IIdiag}
\end{eqnarray}
where for brevity we neglect the index ${\bf k}$, and $\Sigma$ and $G$ are
understood to be functions of $E$. We observe that each term in the integrals
above is a product of a Green's function and a self-energy, i.e., of the form
$G\Sigma$ or $\Sigma G$. These terms can be interpreted as scattering
rates. In particular,  Eq.~(\ref{eq.Idiag}) contains information
about scattering rates into and out of the state $\beta$, and
Eq.~(\ref{eq.IIdiag}) describes scattering into and
out of the state $\alpha$.
To be more specific, we interpret
the terms containing $\Sigma^{\rm ret}$ or 
$\Sigma^{\rm adv}$ as a scattering-out rate, 
e.g, the  term ${\rm Ia}$, $\int dE\,
G^<_{\beta\beta}\,\Sigma^{{\rm adv}(\alpha)}_{\beta\beta}$, can be interpreted
as a rate $\Gamma^{\rm out}_{\beta \rightarrow \alpha}$
for scattering out of state $\beta$ into state $\alpha$.
Similarly,  term ${\rm IIb}$,  $\int dE\,
 \Sigma^{{\rm ret}(\beta)}_{\alpha\alpha}G^<_{\alpha\alpha}$,
is interpreted as a rate 
$\Gamma^{\rm out}_{\alpha \rightarrow \beta}$
for scattering
out of state $\alpha$ into state $\beta$. On the other hand,
the terms containing $\Sigma^<$ are interpreted as scattering-in rates.
Thus, term ${\rm Ib}$ describes the rate 
$\Gamma^{\rm in}_{\alpha \rightarrow \beta}$
for scattering into state $\beta$
from $\alpha$, and term ${\rm IIa}$ describes the rate 
$\Gamma^{\rm in}_{\beta \rightarrow \alpha}$
for scattering into state $\alpha$ from $\beta$. We note here
that $\Gamma^{\rm out}_{\beta \rightarrow \alpha} = 
\Gamma^{\rm in}_{\beta \rightarrow \alpha} < 0$ and 
$\Gamma^{\rm in}_{\alpha \rightarrow \beta} =
\Gamma^{\rm out}_{\alpha \rightarrow \beta} > 0$. 
Combining Eqs.~(\ref{eq.Idiag})
and (\ref{eq.IIdiag}) gives
\begin{eqnarray}
\int dE\; I \,+\, II
& = & \big(\Gamma^{\rm out}_{\beta \rightarrow \alpha}\,+\,
\Gamma^{\rm in}_{\alpha \rightarrow \beta}\big)z_{\beta\beta}
-z_{\alpha\alpha}
\big(
\Gamma^{\rm in}_{\beta \rightarrow \alpha} \,+\,
\Gamma^{\rm out}_{\alpha \rightarrow \beta}
\big)\nonumber \\
& = &
\big(\Gamma^{\rm out/in}_{\beta \rightarrow \alpha}\,+\,
\Gamma^{\rm in/out}_{\alpha \rightarrow \beta}\big) 
(z_{\beta\beta} - z_{\alpha\alpha}).
\end{eqnarray}
Thus, this expression is the product of 
the distance $|z_{\beta\beta} - z_{\alpha\alpha}|$ with
the net transfer rate
($\Gamma^{\rm out/in}_{\beta \rightarrow \alpha}\,+\,
\Gamma^{\rm in/out}_{\alpha \rightarrow \beta}$) between state
$\beta$ and $\alpha$,
and we can interpret this as a velocity or charge transfer rate from e.g.,
$z_{\beta\beta}$ to $z_{\alpha\alpha}$, i.e., a current flow from
$z_{\beta\beta}$ to $z_{\alpha\alpha}$. (The direction of current flow
depends on the net transfer rate.)

%\bibliographystyle{prsty}
%\bibliography{../../all.bib}
%\bibliography{/pub/latex-local/bib/ref.bib}

\begin{thebibliography}{10}
\bibitem{foot.jmf}
We anticipate that $[\hat{H}_{\rm MF},\hat{z}] = 0$ since 
$\hat{H}_{\rm MF}$ depends only on $\hat{z}$. Numerically, we
find $J_{\rm MF} \approx 0$ supporting this expectation. 
Likewise, the numerical result for
$\delta J_{\rm MF}$ in Eq.~(\ref{eq.deljmf}) is negligible.
In Eq.~(\ref{eq.deljmf}), we have neglected 
a term in $[\delta\hat{H}_{\rm MF},\hat{z}] G^{<}$
which is also expected to be zero for the same reason.
\bibitem{foot.ws}
In practice, we diagonalize
$\hat{H}_o + \hat{H}_{\rm MF}$, so that strictly speaking we
obtain a transformation to a basis formed from the eigenstates of 
$\hat{H}_o + \hat{H}_{\rm MF}$, and not the Wannier-Stark basis.
\bibitem{foot.pagendr}
Private communication. H. Page, Thales Research and Technology, Orsay,
France.
The measurements were made with a pulsed
voltage source which is unable to fix a stable voltage
in the NDR region.
\end{thebibliography}

\clearpage
\begin{table}
\begin{ruledtabular}
\begin{tabular}{ccccc}
             & Interface roughness   & LO-phonon    &   Acoustic phonon     \\
	     & \& impurity           &              &                       \\
\hline
intrasubband & $ 0.05$ ps	     &  0.2 ps   &   $> 7$ ps	     \\
intersubband & $\gg 0.2$ ps	     &  $> 0.6$ ps    &   $\gg 8$ ps	     \\
\end{tabular}  
\end{ruledtabular}
\caption{This table gives order-of-magnitude estimates
of inverse scattering rates between Wannier levels
in structure A (see Sec.~\ref{sec.results}) at 77 K. 
The times given
in this table are obtained from  $\hbar/\gamma$ with 
$\gamma$ (in energy units) defined in Eq. (\ref{eq.fgrrate}). More
specifically, for interface roughness and impurity scattering,
$\gamma^{\rm rough}_{(m-n),\mu\nu} + \gamma^{\rm imp}_{\alpha\beta}$ from
Eqs.~(\ref{eq.gamrough}) and (\ref{eq.gamimp}) was used, and for
acoustic phonon scattering, Eq. (\ref{eq.gamac}) is used.
For LO-phonon scattering, Eq. (\ref{eq.gamlo})
without the phonon distribution factor was used.
The parameters used in obtaining these estimates are as follows. For
interface roughness scattering:
$\Delta E_c = 0.27$ eV, $\eta = 0.28$ nm, $\lambda = 10$ nm. For
impurity scattering: $\gamma^{\rm imp}_{\rm par} = 5$ meV. 
For LO-phonon scattering: $E_{\rm lo} = 36.7$~meV, 
$\epsilon_s = 13.1\epsilon_0$, $\epsilon_\infty = 10.9\epsilon_0$.
For acoustic phonon scattering: $\gamma^{\rm ac}_{\rm par} = 0.1$ meV.
\label{tab.rates}
}
\end{table}

\clearpage
\begin{table}
\begin{ruledtabular}
\begin{tabular}{ccccc}
Structure &  $x$  & $n_e$ (cm$^{-2}$)  & $d$ (nm) &  $N_p$ \\
\hline
A (Ref.~\onlinecite{Sir98})  &  0.33 & $3.9\times10^{11}$ & 45.3  &  30    \\
B (Ref.~\onlinecite{Pag01})  &  0.45 & $3.8\times10^{11}$ & 45    &  36    \\
\end{tabular}
\end{ruledtabular}
\caption{Parameters for example GaAs/Al$_x$Ga$_{1-x}$As QC structures:
$x$ is the aluminium content in the barrier, $n_e$ is the 2D carrier
density in one period, $d$ is the period length, and $N_p$ is the
number of periods.
\label{tab.samplepar}
}
\end{table}

\clearpage

\clearpage

\clearpage

\clearpage

\clearpage

\clearpage

\clearpage

\clearpage
\clearpage

\clearpage

\clearpage

\clearpage

\clearpage

\clearpage

\clearpage

\clearpage
\end{document}